\documentclass[
preprint,
superscriptaddress,
 amsmath,amssymb,
 aps,
]{revtex4-1}
\usepackage{subcaption}
\usepackage{xcolor}
\usepackage{graphicx}
\usepackage{dcolumn}
\usepackage{bm}
\usepackage{tikz}
\usepackage[utf8]{inputenc}
\usepackage{amsfonts}
\usepackage{chemfig}
\usepackage{mhchem}
\usetikzlibrary{calc,fadings,decorations.markings}
\usetikzlibrary{decorations.pathreplacing,calc,positioning,shapes.multipart,arrows.meta}
\tikzset{two parallel arrows/.style={decorate,decoration={show path construction,
      lineto code={
       \draw [-latex] ($(\tikzinputsegmentfirst)!#1!90:(\tikzinputsegmentlast)$) 
        -- ($(\tikzinputsegmentlast)!#1!-90:(\tikzinputsegmentfirst)$); 
       \draw [latex-] ($(\tikzinputsegmentfirst)!#1!-90:(\tikzinputsegmentlast)$) 
        -- ($(\tikzinputsegmentlast)!#1!90:(\tikzinputsegmentfirst)$); 
      }}},two parallel arrows/.default=2pt}

\DeclareMathOperator{\sech}{sech}

\begin{document}


\title{Cracking Ion Pairs in the Electrical Double Layer of Ionic Liquids}

\author{Zachary A. H. Goodwin}
\email{zachary.goodwin13@imperial.ac.uk}
\affiliation{Department of Chemistry, Imperial College of London, Molecular Sciences Research Hub, White City Campus, Wood Lane, London W12 0BZ, UK}
\affiliation{Thomas Young Centre for Theory and Simulation of Materials, Imperial College of London, South Kensington Campus, London SW7 2AZ, UK}

\author{Alexei A. Kornyshev}
\email{a.kornyshev@imperial.ac.uk}
\affiliation{Department of Chemistry, Imperial College of London, Molecular Sciences Research Hub, White City Campus, Wood Lane, London W12 0BZ, UK}
\affiliation{Thomas Young Centre for Theory and Simulation of Materials, Imperial College of London, South Kensington Campus, London SW7 2AZ, UK}
\affiliation{Institute of Molecular Science and Engineering, Imperial College of London, South Kensington Campus, London SW7 2AZ, UK}


\date{\today}
             

\begin{abstract}
Here we investigate a limiting case of the theory for aggregation and gelation in the electrical double layer (EDL) of ionic liquids (ILs). The limiting case investigated only accounts for ion pairs, ignoring the possibility of larger clusters and a percolating ionic network. This simplification, however, permits analytical solutions for the properties of the EDL. The resulting equations demonstrate the competition between the free energy of an association and the electrostatic potential in the EDL. For small electrostatic potentials and large negative free energies of associations, the ion pairs dominate in the EDL. Whereas, for electrostatic potential energies larger than the free energy of an association, electric-field-induced cracking of ion pairs occurs. The differential capacitance for this consistent ion pairing theory has a propensity to have a ``double hump camel'' shape. We compare this theory against previous free ion approaches, which do not consistently treat the reversible associations in the EDL.
\end{abstract}
   
\maketitle

\section{Introduction}

Ionic liquids (ILs), an electrolyte solely composed of molecular cations and anions, are of interest for applications in energy storage devices because of their ability to withstand larger voltages without decomposing than aqueous electrolytes~\cite{Welton1999,Hermann2008,Hallett2011,Kondrat2016,son2020ion,Fedorov2014}. This has motivated many to study the electrical double layer (EDL) of ILs~\cite{Fedorov2015Re,Trulsson2010,Sha2014,Vatamanu2012,Merlet2014,Merlet2013,Merlet2011,bhuiyan2012monte,Lamperski2014,Forsman2011,gavish2016,Han2014,Maggs2016,Girotto2017,Limmer2015,Lauw2009}, i.e., how the ions arrange at a charged interface~\cite{Levin2002,Fedorov2014,goodwin2021review}. This interest was further intensified upon the discovery, from surface-force balance measurements, that the EDL of ILs have extremely long-ranged monotonic interactions~\cite{Gebbie2013,Gebbie2015,Smith2016,Gebbie2017rev,smith2017struct,Han2020IL,Jurado2016,Jurado2015,Jurado2017EDL,Mao2019nano,Hjalmarsson2017,comtet2017nano}, in contrast to the decaying oscillations of charge (overscreening) which is expected to occur in such concentrated electrolytes~\cite{Fedorov2008a,Fedorov2008,Georgi2010,Bazant2011,Coles2020length,pedroRTILs,pedro2022force,gavish2018solvent,emily2021,levy2019spin,gavish2018solvent}. This was interpreted by Gebbie \textit{et al.}~\cite{Gebbie2013,Gebbie2015} as ILs behaving as dilute electrolytes, with over 99.99\% of the ions bound up in neutral ion pairs in the bulk. This raised the questions of how many ions were paired in ILs~\cite{Zhao2009,Lee2015,Zhang2015,Kirchner2014,Araque2015,Krichner2015ionpair}, and how these ion pairs are distributed in the EDL of ILs~\cite{Ma2015,adar2017bjerrum,yufan2020}?

One estimate for the number of ion pairs in bulk ILs was provided by Lee \textit{et al.}~\cite{Lee2015}, where it was found that 2/3 of ions were free. This estimate was based on a Debye-H\"{u}ckel theory coupled to the mass action law of ion pair formation, where the equilibrium constant is controlled by the potential energy based on the screened interaction. As ions in ILs are very densely packed, the formally calculated Debye screening length comes out shorter than an ion diameter, which means the ions effectively did not interact in that theory~\cite{Lee2015}, and the entropy of mixing determines the proportion of free ions. On the other hand, Feng \textit{et al.}~\cite{feng2019free} utilised a dynamical criteria and molecular dynamics simulation of typical ILs to find that just 10-20\% of ions can be considered, on average, free. See Ref.~\citenum{Krichner2015ionpair} for a review on the extent of ion pair formation in ILs. While the quantitative extent of ion pairing varies between studies, all of them lead to the conclusion that the fraction of free ions is significantly higher than that suggested by Gebbie \textit{et al}~\cite{Gebbie2013,Gebbie2015}.

In the EDL, Ma \textit{et al.}~\cite{Ma2015} developed a sophisticated classical density functional theory for ILs with free ions and ion pairs. It was found that the charge density profiles were not too sensitive to the extent of ion pairing, suggesting a connection between ion pairs and overscreening~\cite{Ma2015}. Furthermore, Avni \textit{et al.}~\cite{avni2020charge} established a link between ion pairs (and small aggregates) and overscreening equations~\cite{Bazant2011} in the long wavelength limit. On a more qualitative level, Ref.~\citenum{Chen2017} developed a local density approach based on the fraction of free ions, where a transition from a camel to bell shaped differential capacitance curve was predicted from increasing the temperature~\cite{Chen2017,goodwin2017mean,goodwin2017underscreening}. Moreover, Zhang \textit{et al.}~\cite{yufan2020} developed a semi-phenomenological theory that included dielectric saturation of the IL from ion pair orientation, as well as their expulsion from the double layer.

In such a concentrated system as ILs, one does not expect formation of only ion pairs, but not also larger clusters of ions~\cite{Dupont2004,Singh2008,Dupont2011}. In molecular dynamics simulations, it has been shown that indeed larger aggregates, percolating ionic networks and heterogeneities can form~\cite{Wang2005,Hu2006hetero,Bernardes2011,Lopes2006,borodin2017liquid,choi2018graph,jeon2020modeling}. Recently, McEldrew \textit{et al.}~\cite{mceldrew2020theory,mceldrew2020correlated,mceldrew2021wise,mceldrew2021salt} developed a description of ILs in the bulk, based on the works of Flory~\cite{flory1941molecular,flory1941molecular2,flory1942thermodynamics,flory1942constitution,flory1953principles}, Stockmayer~\cite{stockmayer1943theory,stockmayer1944theory} and Tanaka~\cite{tanaka1989,tanaka1990thermodynamic,tanaka1994,tanaka1995,ishida1997,tanaka1998,tanaka1999,tanaka2002} in polymer physics, for the formation of ionic aggregates and a percolating ionic network, i.e. a gel~\cite{Reber2020}. This was applied to ILs~\cite{mceldrew2020correlated}, and other super-concentrated electrolytes~\cite{mceldrew2021wise,mceldrew2021salt}, and a consistent theory of ionic transport based on vehicular motion of clusters was also developed~\cite{mceldrew2020correlated,france2019}. 

In Ref.~\citenum{Goodwin2022EDLgel}, a consistent treatment of ionic associations in the EDL and bulk~\cite{mceldrew2020theory}, and the equilibrium between them was established with a Boltzmann closure relation of the free ions. When free ions dominate, little difference was found with previous free ion theories~\cite{Goodwin2022EDLgel}. However, when the system has a significant number of associations, it was found that the gel can screen electrode charge, owing to the reversible equilibrium with unequal numbers of cations and anions forming a charged gel~\cite{Goodwin2022EDLgel}. Moreover, the gel dominates screening at linear response, and the gelation-crowding transition~\cite{Goodwin2022EDLgel} was found to be conceptually similar to the well-known overscreening-crowding transition~\cite{Bazant2011}. The theory in Ref.~\citenum{Goodwin2022EDLgel} was only solved numerically when each ions could form a maximum of 3 associations with co-ions (or 4 when comparing against experiments). 

\begin{figure}
    \centering
    \includegraphics[width=0.45\textwidth]{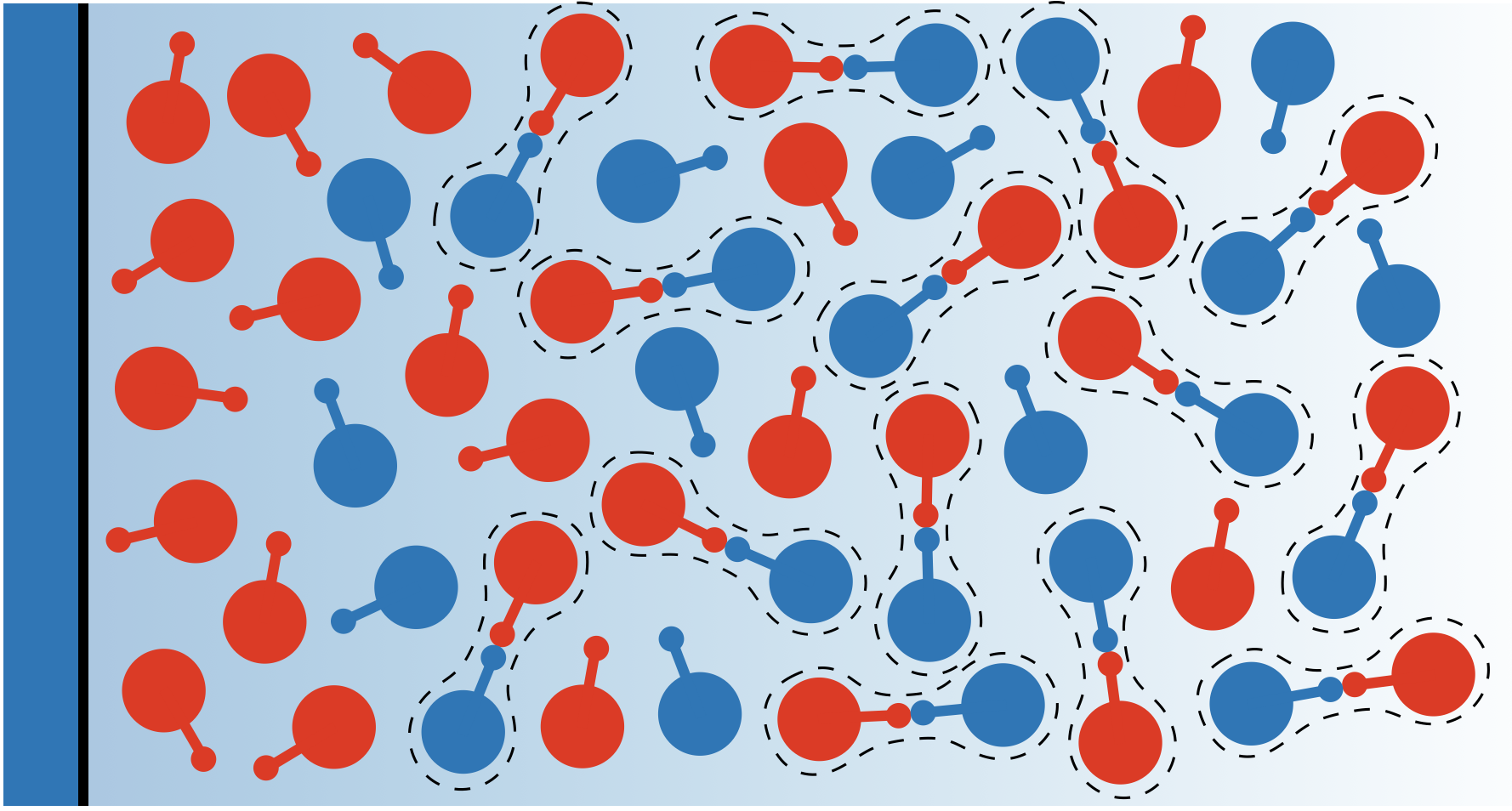}
    \caption{\textbf{Schematic of ion pairs being cracked in the electrical double layer of an ionic liquid}. Cations are shown in red, and anions in blue, with a negatively charged interface on the left. Both cations and anions have a functionality of 1, i.e., they can form one bond. Ion pairs are shown by the dangling bond of a cation and anion touching, and the dotted line enclosing the pair of ions. Close to the charged interface, ion pairs are cracked in favour of free cations.}
    \label{fig:IP_EDL}
\end{figure}

Here we study a limiting case of that theory, where we only account for the formation of ion pairs. While McEldrew \textit{et al.}~\cite{mceldrew2020correlated} have shown that ions in ILs can form between 3 and 5 associations, studying the limit of ion pair formation is conceptually simple, and ILs which only form ion pairs could, perhaps, be found. In Fig.~\ref{fig:IP_EDL} we schematically show the qualitative picture of the presented theory. In the bulk, ion pairs co-exist with free cations and anions, with a reversible equilibrium between them. In the EDL at a negatively charged interface, free cations are favoured, which causes an accumulation of free cations. Here we clearly demonstrate that there is an electric-field-induced cracking of ion pairs, which gives rise to a higher propensity to a ``double hump camel'' differential capacitance curve. 

The paper is structured as follows. First, the bulk theory of ion pairs is outlined, which is similar to that outlined in Refs.~\citenum{mceldrew2020theory,mceldrew2020correlated,mceldrew2021salt,mceldrew2021wise,Goodwin2022EDLgel}. Next, the Bolztmann closure relation for ion pairs is solved exactly~\cite{Goodwin2022EDLgel}, which permits the properties of the EDL to be analytically investigated. The presented theory is then compared against the free ion theory of Ref.~\citenum{Chen2017} \textcolor{black}{as well as with pertinent experimental differential capacitance data}. Finally, we discuss this theory in the context of other approaches, highlighting its limitations and possible extensions.

\section{Theory}

\subsection{Ion pairing}

Here we take the limit of Refs.~\citenum{mceldrew2020correlated,Goodwin2022EDLgel} of a symmetric, incompressible IL that can only form ion pairs. The incompressibility condition means no voids are accounted for, and it is symmetric because the volumes of cations and anions are the same ($v_+ = v_- = v$), and the number of associations cations and anions can form are both 1 (this is referred to as the functionality, given by $f_+ = f_- = f = 1$). The limit of ion pairs, where 1 cation can only bind to 1 anion, is studied here as analytical solutions are obtainable, which provides insight into how the associations are destroyed by electrostatic potentials. For more information about this theory beyond ion pair formation, the reader is referred to Refs.~\citenum{mceldrew2020theory,mceldrew2020correlated}.

The theory is based on a lattice-gas model~\cite{mceldrew2020theory,mceldrew2020correlated}. The total number of lattice sites is given by
\begin{equation}
\Omega = \sum_{lm}(l + m)N_{lm},
\end{equation}

\noindent where $N_{lm}$ is the number of species of rank $lm$ (either free cations $10$, free anions $01$, or ion pairs $11$). Note no gel can form as clusters are limited to ion pairs here, i.e. the number of cations $l$ is limited to 1 and the number of anions $m$ is also limited to 1. Dividing by the total number of lattice sites yields
\begin{equation}
1 = \sum_{lm}(l + m)c_{lm},
\end{equation}

\noindent where $c_{lm} = N_{lm}/\Omega$ is the dimensionless concentration ($\#$ per lattice site) of species of rank $lm$. The volume fraction of a species of rank $lm$ is given by $\phi_{lm} = (l + m)c_{lm}$. The volume fraction of cations/anions is 
\begin{equation}
    \phi_{+/-} = \sum_{lm}l/m~c_{lm}.
\end{equation}

The free energy of the mixture~\cite{mceldrew2020theory,mceldrew2020correlated} is taken to be
\begin{align}
\beta \mathcal{F} = \sum_{lm} N_{lm}\ln \left( \phi_{lm} \right)+N_{lm}\Delta_{lm}
\label{eq:Fb}
\end{align}

\noindent where $\beta = 1/k_BT$ is inverse thermal energy, $\Delta_{lm}$ is the free energy of formation of a species of rank $lm$ from free cations and anions, measured in units of $k_BT$. Further details of the free energy of formation of species can be found in Refs.~\citenum{mceldrew2020theory,mceldrew2020correlated}. 

As shown in detail in Refs.~\citenum{mceldrew2020theory,mceldrew2020correlated,mceldrew2021salt}, the ion pair equilibrium is expressed as
\begin{equation}
    c_{11} = \lambda \phi_{10} \phi_{01}.
\end{equation}

\noindent where $\lambda = \exp(-\beta \Delta f_{+-})$ is the ionic association constant. The free energy of an association is given by $\Delta f_{+-} = \Delta u_{+-} - T\Delta s_{+-}$, where the binding energy is $\Delta u_{+-}$, and entropy of an association is $\Delta s_{+-}$. Changing the association constant is equivalent to changing the temperature: a larger association constant is equivalent to a lower temperature, and a smaller one is at higher temperatures. The reader is referred to Refs.~\citenum{mceldrew2020correlated,Goodwin2022EDLgel} for further details about the association constant and temperature dependence. 

The ion pair equilibrium is expressed in terms of free cations, $\phi_{10}$, and free anions, $\phi_{01}$. These volume fractions are related to the volume fractions of cations and anions through $\phi_{10} = \phi_+(1 - p_{+-})$ and $\phi_{01} = \phi_-(1 - p_{-+})$, respectively, where $p_{ij}$ is the probability that an association site of species $i$ is bound to species $j$, has been introduced~\cite{mceldrew2020theory}. 

The number of associations per lattice site is given by
\begin{equation}
    \zeta = \phi_+p_{+-} = \phi_-p_{-+},
    \label{eq:p1}
\end{equation}

\noindent which is also a statement of the conservation of associations. The mass action law of the association equilibrium is given by
\begin{equation}
    \lambda\zeta = \frac{p_{+-}p_{-+}}{(1-p_{+-})(1-p_{-+})}.
    \label{eq:p2}
\end{equation}

\noindent These two equations can be explicitly solved for the association probabilities, or equivalently the number of associations per lattice site
\begin{equation}
    \zeta = \dfrac{1 + \lambda - \sqrt{1 + 2\lambda + \lambda^2(1 - 2\phi_{+/-})^2}}{2\lambda}.
\label{eq:zeta_bulk}
\end{equation}

\subsection{Boltzmann closure}

In Ref.~\citenum{Goodwin2022EDLgel}, a Boltzmann closure relation was introduced to ensure the chemical equilibria was consistently held in the EDL and the bulk. In the limit of ion pair formation, this closure relation is given by
\begin{equation}
   e^{-2\alpha u} = \dfrac{\bar{\phi}_+(1 - \bar{p}_{+-})}{\bar{\phi}_-(1 - \bar{p}_{-+})},
\end{equation}

\noindent where $u$ is the electrostatic potential in units of \textcolor{black}{thermal voltage}, and $\alpha$ is a parameter which represents short-range repulsion between ions of the same sign beyond mean-field (see Refs.~\citenum{goodwin2017mean,BBKGK} for more details). Note a bar is used to denote quantities which are in the EDL.




This closure relation can be explicitly solved for the volume fractions of cations (or anions )
\begin{equation}
    \bar{\phi}_{+} = \dfrac{1}{2} + \dfrac{\sinh (2\alpha u)}{4\lambda}\left\{1 - \sqrt{1 + \dfrac{4\lambda}{1 + \cosh (2\alpha u)}}\right\}.
    \label{eq:EDL_phip}
\end{equation}

\noindent The anion volume fraction can be obtained from the incompressibility constraint, $\bar{\phi}_- = 1 - \bar{\phi}_+$. In the limit of vanishing $\lambda$, and therefore associations, the equation reduces to $\bar{\phi}_{+} = [1 - \tanh(\alpha u)]/2$, which is the expected limit of an incompressible IL. In the limit of large fields, the equation again reduces to $\bar{\phi}_{+} = [1 - \tanh(\alpha u)]/2$, which demonstrates the dissociation of ion pairs in electrostatic potentials. Therefore, Eq.~\eqref{eq:EDL_phip} is a Fermi-type function (it shall be shown later).

\begin{figure}
    \centering
    \includegraphics[width=0.45\textwidth]{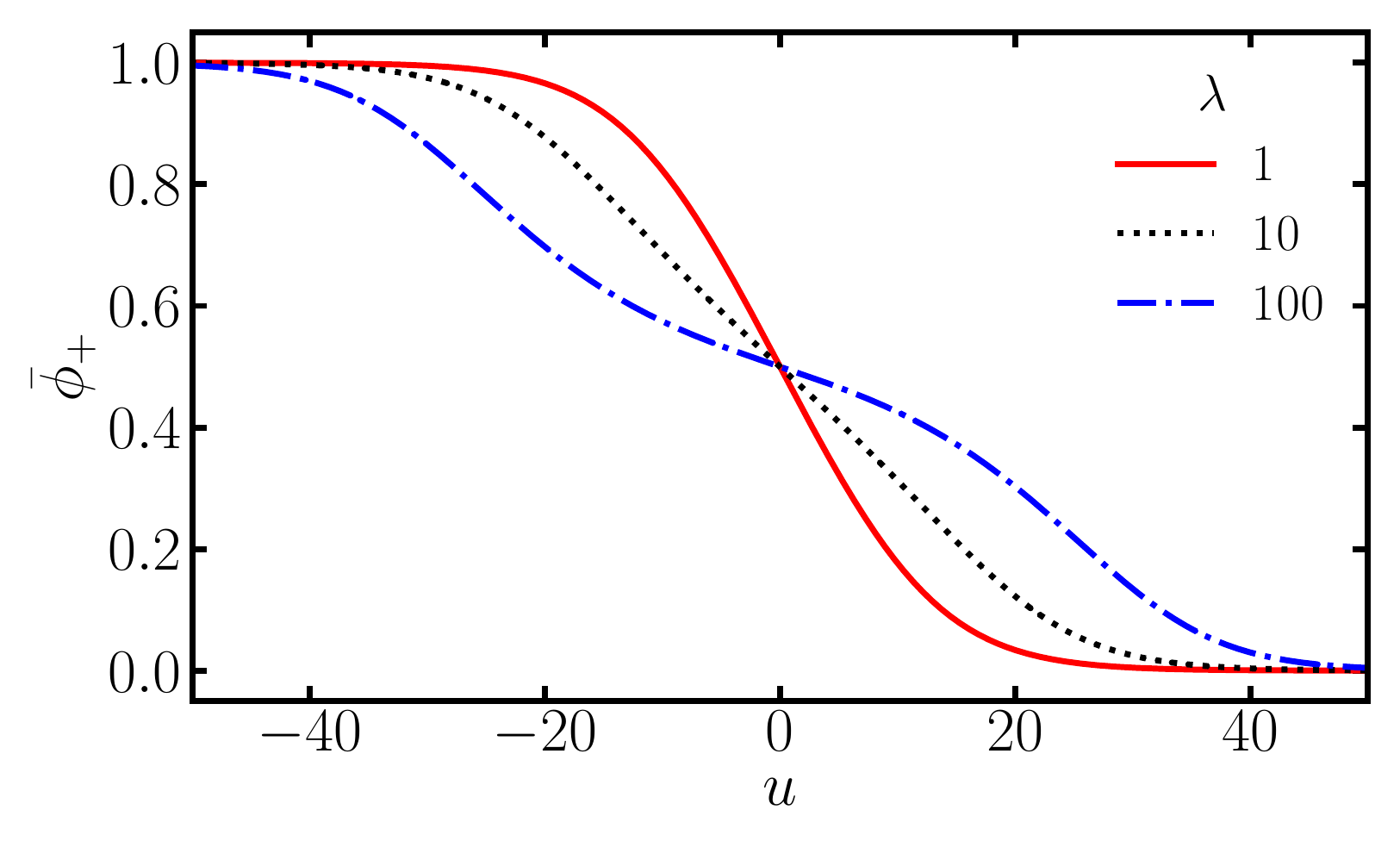}
    \caption{\textbf{Large association constants have a clear energy scale at which ion pairs dissociate}. Volume fraction of cations as a function of electrostatic potential, in units of \textcolor{black}{thermal voltage}, for the indicated association constants. Here $\alpha = 0.1$.}
    \label{fig:phi}
\end{figure}

A notable feature of Eq.~\eqref{eq:EDL_phip} is that the hyperbolic functions always appear as a ratio with the association constant. Recall, that the association constant is the exponential of minus the free energy of an association. Therefore, there is a clear competition between the energy scale of an association and the electrostatic potential. When $-\beta\Delta f_{+-} > 2\alpha|u|$, the system is in the limit of associations dominating over the electrostatic potential. Whereas, for $2\alpha|u| > -\beta\Delta f_{+-}$, the electrostatic potential dominates over the free energy of associations, and free ions prevail. 

In Fig.~\ref{fig:phi}, Eq.~\eqref{eq:EDL_phip} is plotted as a function of $u$ for several association constants (for anions, these curves are reflected at $u=0$). For $\lambda = 1$, the fraction of free ions is 0.73, and $\bar{\phi}_+$ resembles a Fermi function. A similar dependence of the volume fraction of cations on $u$ is observed for $\lambda = 10$, where 36\% of ions are free. Finally, for $\lambda = 100$, with a free ion fraction of 0.13, $\bar{\phi}_+$ has a more complicated dependence on $u$. Initially, $\bar{\phi}_+$ changes linearly with a small slope, but after $u \approx \pm20$, there is a rapid change in the volume fraction of cations to 0 or 1. This transition is approximately where $\ln(\lambda) = -\beta\Delta f_{+-} \approx 2\alpha|u|$, which demonstrates that for larger electrostatic potentials the ion pairs are destroyed in favour of free ions.

Furthermore, Eq.~\eqref{eq:EDL_phip} can be inserted into the expression for the number of associations per lattice site, to obtain
\begin{equation}
  \bar{\zeta} = \dfrac{1 + \lambda - \sqrt{1 + 2\lambda + \left(\dfrac{\sinh (2\alpha u)}{2}\left\{1 - \sqrt{1 + \dfrac{4\lambda}{1 + \cosh (2\alpha u)}}\right\}\right)^2}}{2\lambda}.
\label{eq:EDL_zeta}
\end{equation}

\begin{figure}
    \centering
    \includegraphics[width=0.45\textwidth]{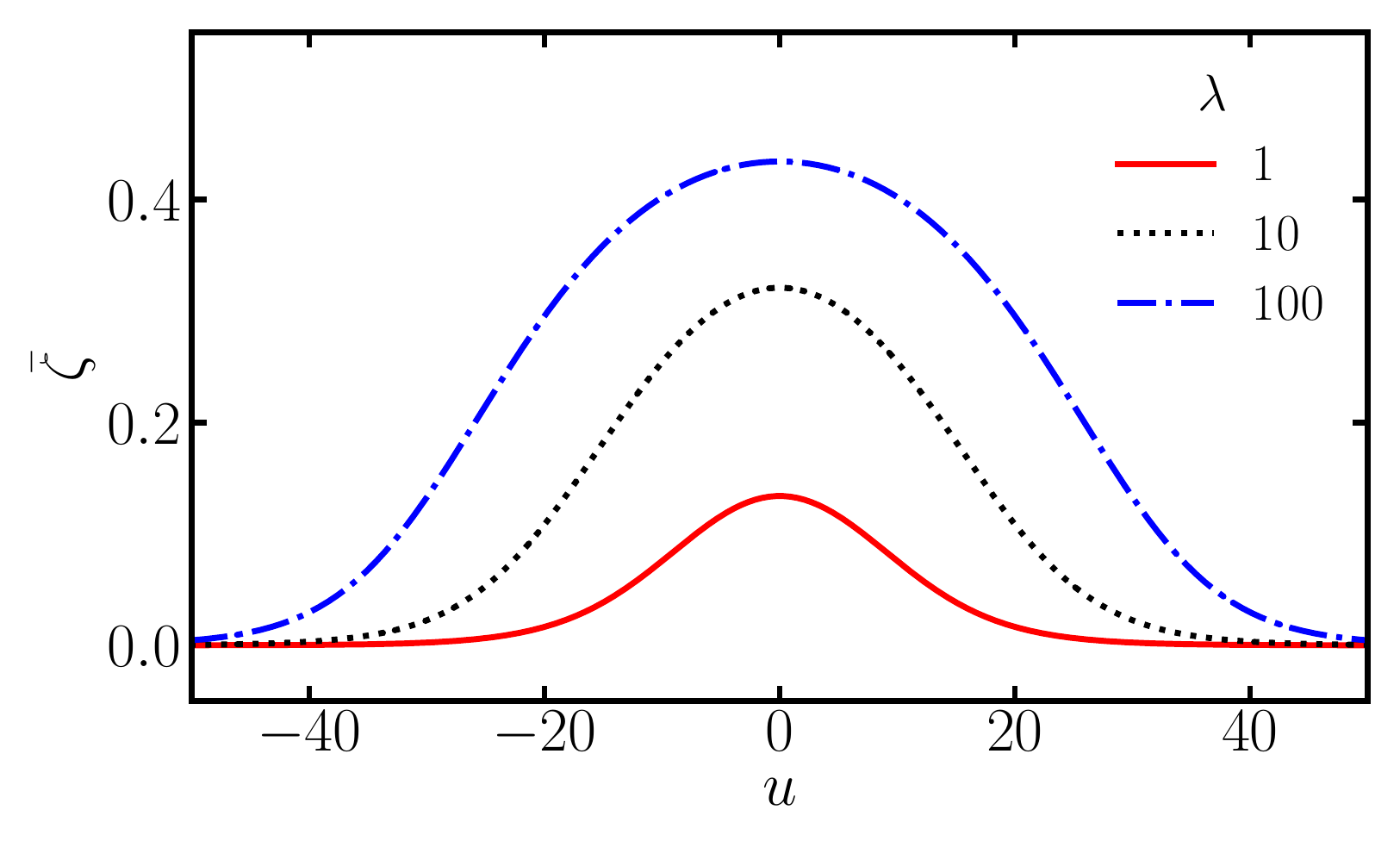}
    \caption{\textbf{Electric-field-induced destruction of associations}. Number of associations per lattice site as a function of electrode potential, in units of \textcolor{black}{thermal voltage}, for the indicated association constants. Here $\alpha = 0.1$.}
    \label{fig:zeta}
\end{figure}

\noindent In the limit of large fields, this reduces to $\bar{\zeta} = \left(1 + \lambda\right)\left[1 - |\tanh(\alpha u)|\right]/2\lambda$, and therefore, $\bar{\zeta}$ tends to 0 at large electrostatic potentials. This function is plotted in Fig.~\ref{fig:zeta}, again for different values of the association constant. For all values of $\lambda$, $\bar{\zeta}$ reaches 0 at large electrostatic potentials. Near the potential of zero charge, $\bar{\zeta}$ is at a maximum, with values closer to $1/2$ for larger $\lambda$.

The modified Poisson-Fermi equation is given by
\begin{equation}
    \nabla^2u = \kappa^2\dfrac{\sinh (2\alpha u)}{2\lambda}\left\{\sqrt{1 + \dfrac{4\lambda}{1 + \cosh (2\alpha u)}}-1\right\},
\end{equation}

\noindent where the inverse Debye length is $\kappa = \sqrt{v\epsilon_0\epsilon/e^2\beta}$. At linear response, we obtain
\begin{equation}
    \nabla^2u = \kappa^2(1-p)\alpha u,
\end{equation}

\noindent which is the expected limit from Ref.~\citenum{Goodwin2022EDLgel}, where the screening length~\cite{adar2017bjerrum,IBL} is given by
\begin{equation}
    \ell = \dfrac{1}{\kappa\sqrt{\alpha(1-p)}}.
\end{equation}

\noindent Here $p = (1 + \lambda - \sqrt{1 + 2\lambda})/\lambda$ is the association probability in the bulk, where $\phi_+ = \phi_- = 1/2$. 

Converting to dimensionless units, $\kappa\nabla = \tilde{\nabla}$, and taking the first integral, we obtain
\begin{multline}
\alpha(\tilde{\nabla} u)^2 = \ln\left\{\dfrac{\cosh(\alpha u)\left[1 + 2\lambda + \sqrt{1 + 2\lambda \sech^2(\alpha u)}\right] - 2\lambda\sinh(\alpha u)}{1 + 2\lambda + \sqrt{1 + 2\lambda}} \right\} + \\ \ln\left\{\dfrac{\cosh(\alpha u)\left[1 + 2\lambda + \sqrt{1 + 2\lambda \sech^2(\alpha u)}\right] + 2\lambda\sinh(\alpha u)}{1 + 2\lambda + \sqrt{1 + 2\lambda}} \right\} + \\ \dfrac{\cosh^2(\alpha u)\left[\sqrt{1 + 2\lambda \sech^2(\alpha u)} - 1\right] + 1 - \sqrt{1 + 2\lambda}}{\lambda}.
\label{eq:sigma_new}
\end{multline}

\noindent The square root of this function yields the dimensionless surface charge density. In the limit of vanishing $\lambda$, we obtain $(\tilde{\nabla} u)^2 = 2 \ln\left\{\cosh(\alpha u)\right\}/\alpha$, which is the expected limit. Moreover, in the limit of large electrostatic potentials, we obtain $|\tilde{\nabla} u| = \sqrt{|u|}$, which is again the expected result.



\begin{figure}
    \centering
    \includegraphics[width=0.45\textwidth]{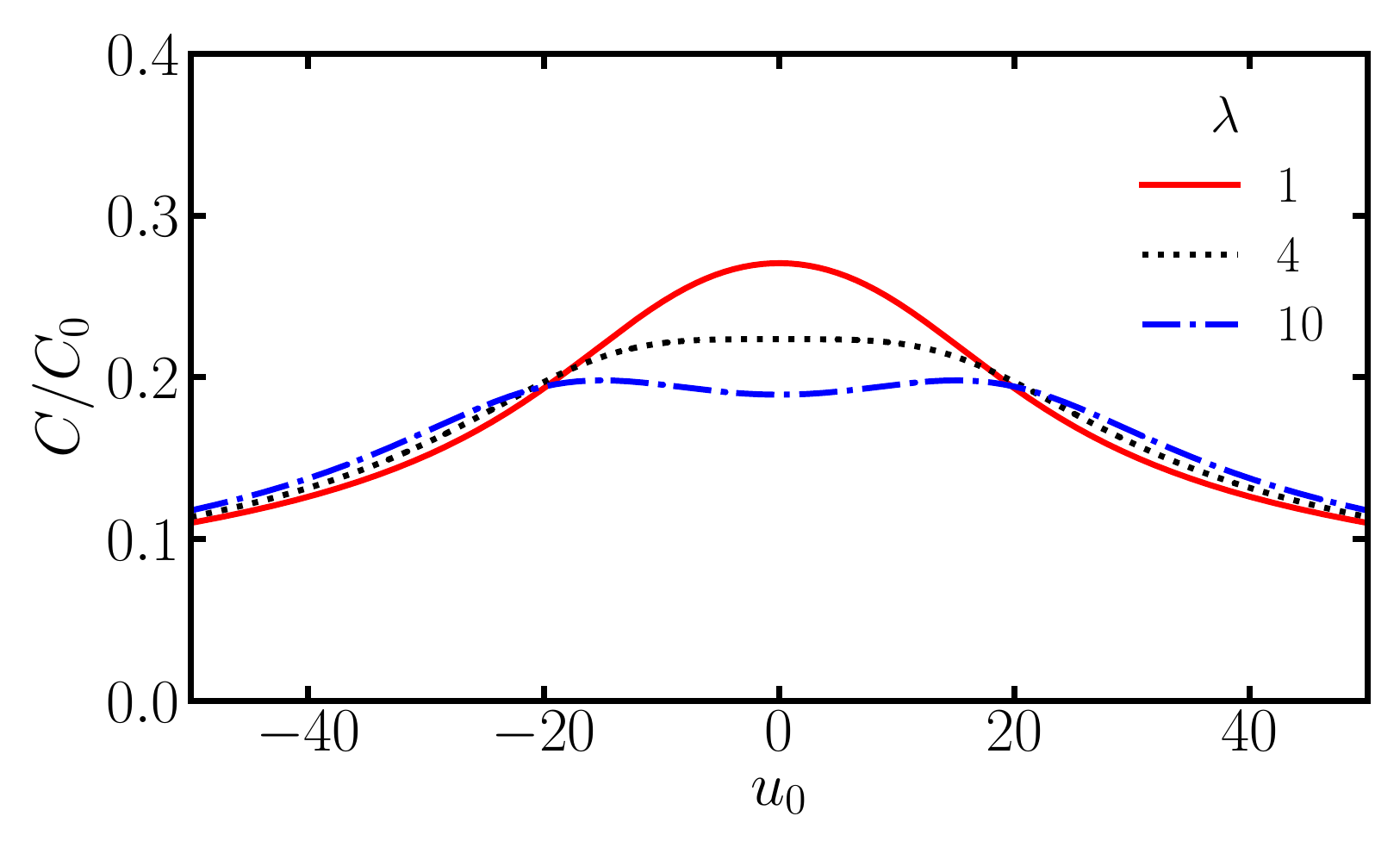}
    \caption{\textbf{The ``camel-to-bell'' transition occurs at $\lambda = 4$ with 1/2 of free ions}. Differential capacitance, in units of Debye capacitance, as a function of electrostatic potential drop across the entire electrical double layer, in units of \textcolor{black}{thermal voltage}. The differential capacitance is shown for three association constants, one with a bell shape ($\lambda = 1$), one at the transition between bell and camel ($\lambda = 4$), and another camel shaped curve ($\lambda = 10$). Here $\alpha = 0.1$ and Debye capacitance is $C_0 \approx 75-100~\mu$Fcm$^{-2}$.}
    \label{fig:DC}
\end{figure}

Taking the derivative of the surface charge density with respect to the potential drop across the EDL, $u_0$, yields the differential capacitance
\begin{equation}
    C = C_0\dfrac{d \tilde{\sigma}}{d u_0},
\end{equation}

\noindent where $C_0 = \epsilon\epsilon_0\kappa$ is the Debye capacitance, and the dimensionless surface charge density is $\tilde{\sigma} = \pm\sqrt{(\tilde{\nabla} u)^2}$. Taking the derivative, we obtain






\begin{equation}
\dfrac{C}{C_0} = \sqrt{\alpha}\dfrac{2(1 + 2\lambda)|\tanh(\alpha u_0)|\left\{1 + \sqrt{1 + 2\lambda \sech^2(\alpha u_0)}\right\}}{\left[\left(1 + 2\lambda + \sqrt{1 + 2\lambda \sech^2(\alpha u_0)}\right)^2 - 4\lambda^2\tanh^2(\alpha u_0) \right]\sqrt{\alpha(\tilde{\nabla} u_0)^2}},
\end{equation}

\noindent where Eq.~\eqref{eq:sigma_new} must be utilised for the term in the denominator. In the limit of vanishing $\lambda$, the equation reduces to $C/C_0 = \sqrt{\alpha}|\tanh(\alpha u_0)|/\sqrt{2\ln\{\cosh(\alpha u_0)\}}$, which is the expected limit of an incompressible IL without associations~\cite{Kornyshev2007,kilic2007a,Bazant2009a,Fedorov2014,Chen2017,goodwin2017mean}. In the limit of large $\lambda$, when there are very few free ions, it is expected that this equation reduces to that of Gouy-Chapman, although this has not been explicitly obtained.

At large electrostatic potentials, the differential capacitance reduces to $C/C_0 = 1/\sqrt{2|u_0|}$, which is the same relationship as the charge conservation law obtained by Kornyshev~\cite{Kornyshev2007}. In the limit of small potentials, the differential capacitance can be expanded as a power series to obtain
\begin{equation}
    \dfrac{C}{C_0} = \sqrt{\alpha}\left[\sqrt{1 - p} + \dfrac{\sqrt{1 + 2\lambda} - 3}{4\sqrt{2(1 + 2\lambda)(1 + \sqrt{1 + 2\lambda})}}(\alpha u_0)^2 \right].
    \label{eq:DC_New}
\end{equation}

\noindent Clearly, the capacitance at zero charge is $C/C_0 = \sqrt{\alpha(1-p)}$, which is consistent with previous free ion approaches~\cite{Chen2017,adar2017bjerrum,IBL}. However, the sign of the quadratic term changes when $\lambda = 4$, which corresponds to $1/2$ of ions free in the bulk. When there are more than $1/2$ of free ions, a ``bell''-shaped differential capacitance is obtained; while for fewer free ions than $1/2$, there is a ``double hump camel''-shape. In Fig.~\ref{fig:DC}, Eq.~\eqref{eq:DC_New} is plotted as a function of potential drop across the entire EDL, for several association constants. For $\lambda = 1$, there is a clear ``bell'' shape and for $\lambda = 10$ there is a clear ``double hump camel'' shape, but for $\lambda = 4$ the differential capacitance is very flat, as it is right at the ``camel-to-bell'' transition of the presented theory.

\begin{figure}
    \centering
    \includegraphics[width=0.45\textwidth]{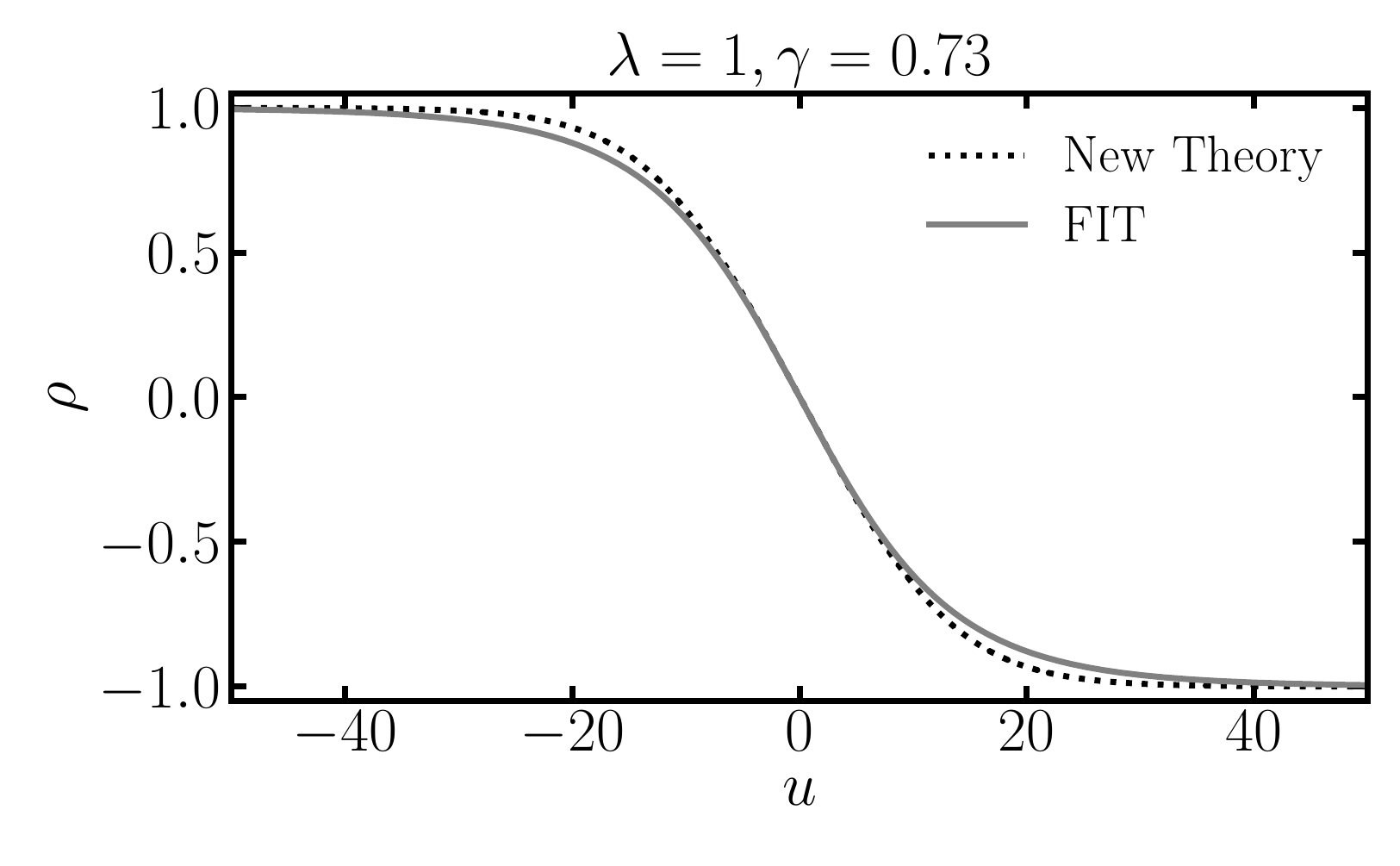}
    \centering
    \includegraphics[width=0.45\textwidth]{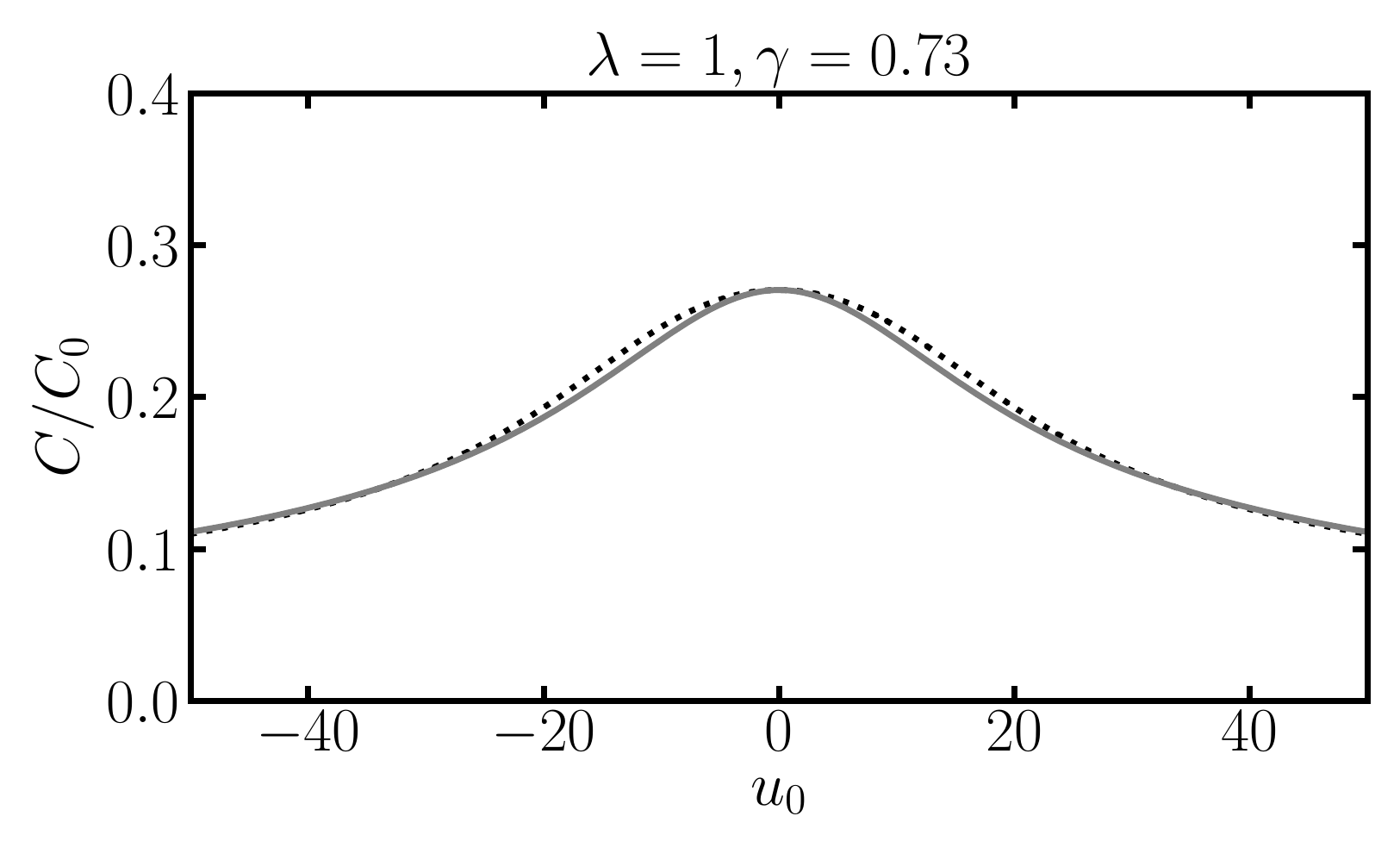}
    \centering
    \includegraphics[width=0.45\textwidth]{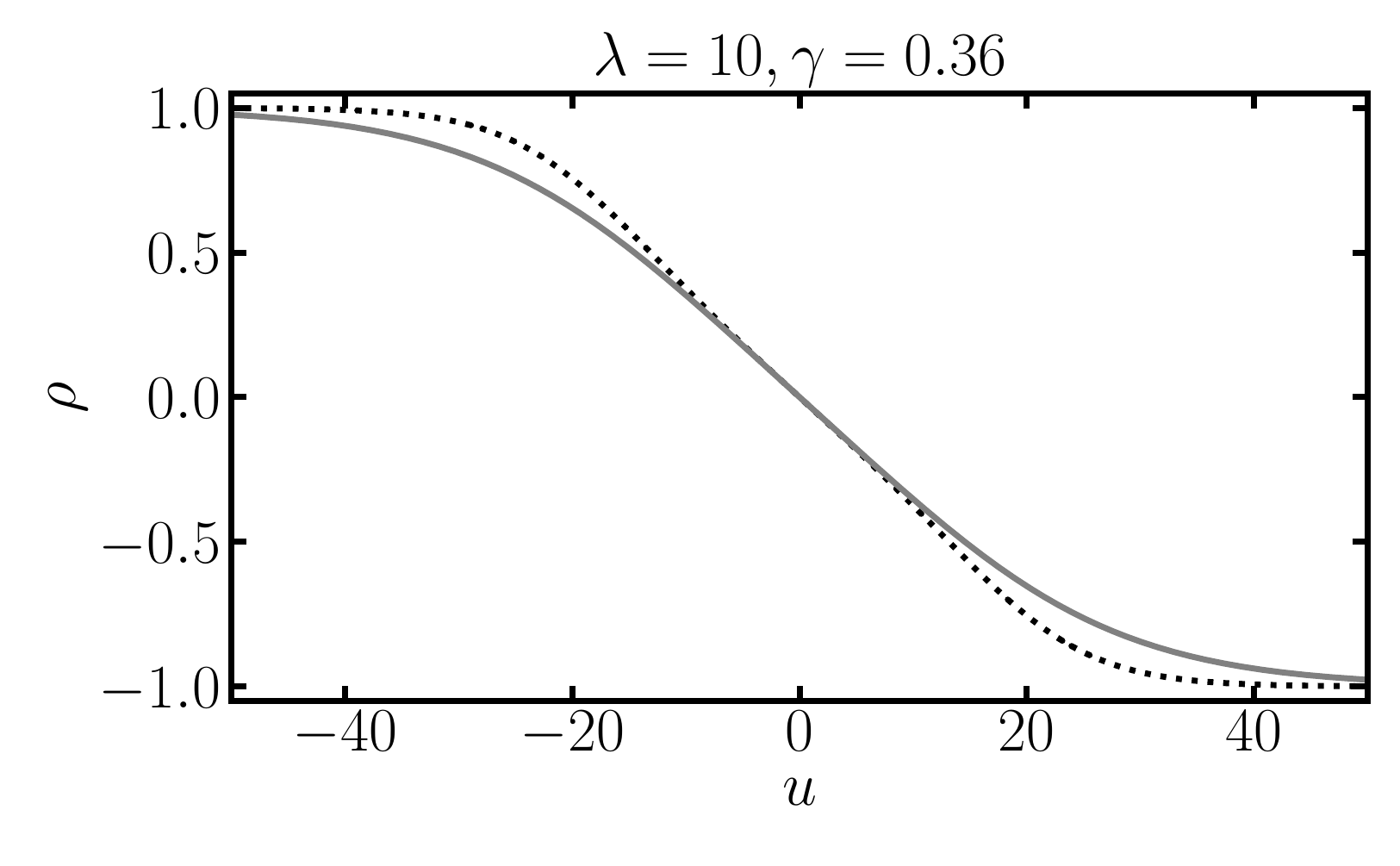}
    \centering
    \includegraphics[width=0.45\textwidth]{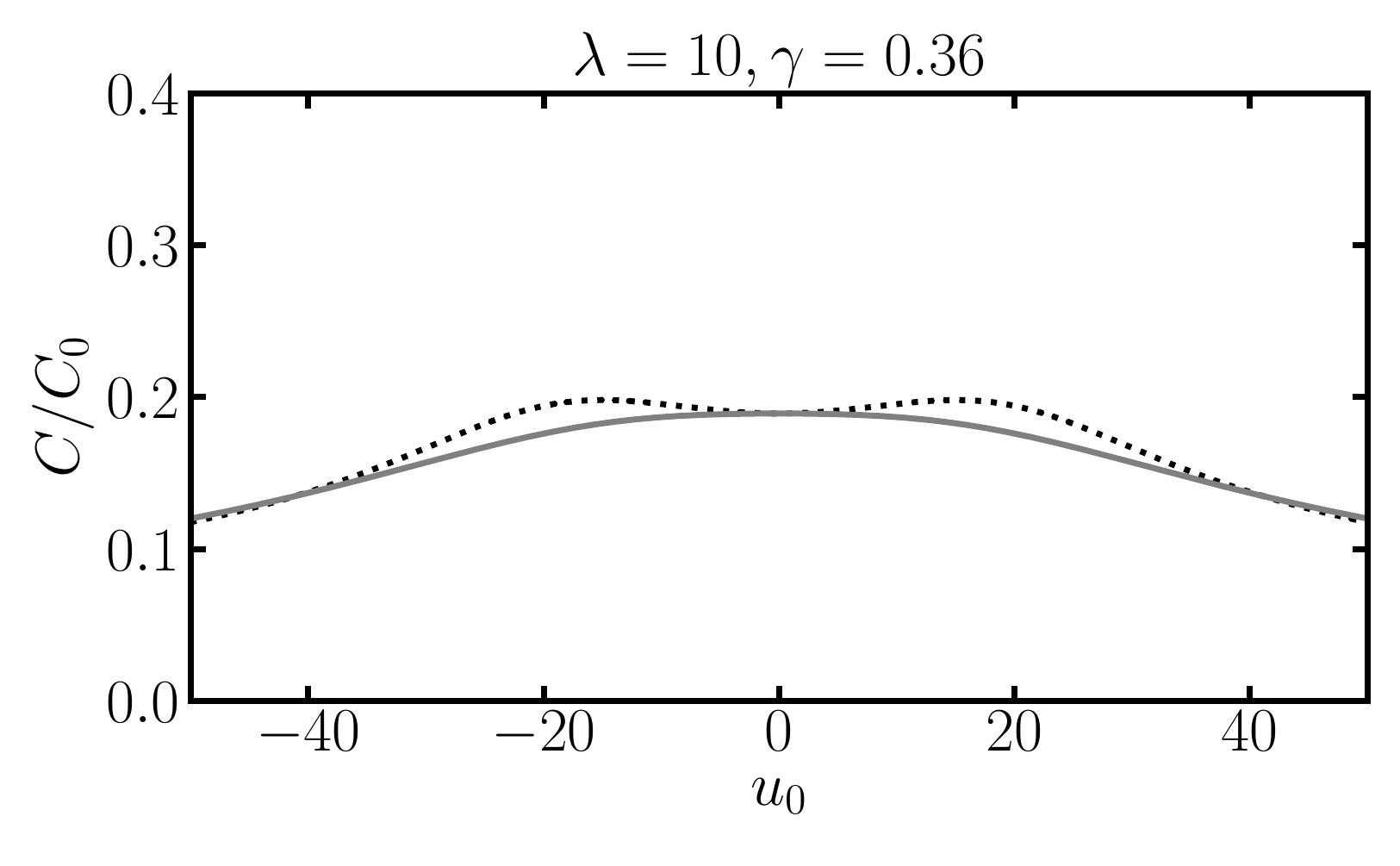}
    \centering
    \includegraphics[width=0.45\textwidth]{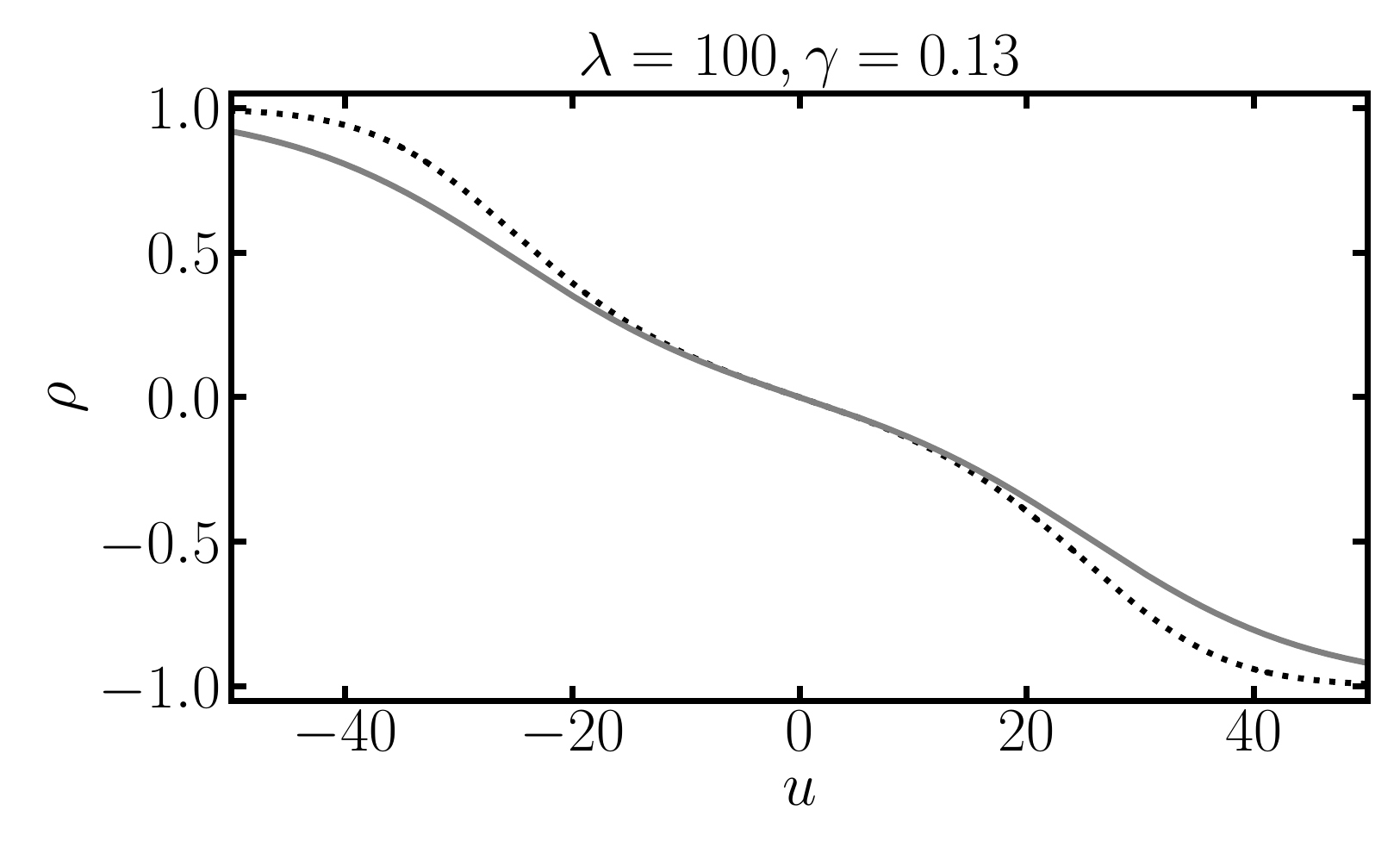}
    \centering
    \includegraphics[width=0.45\textwidth]{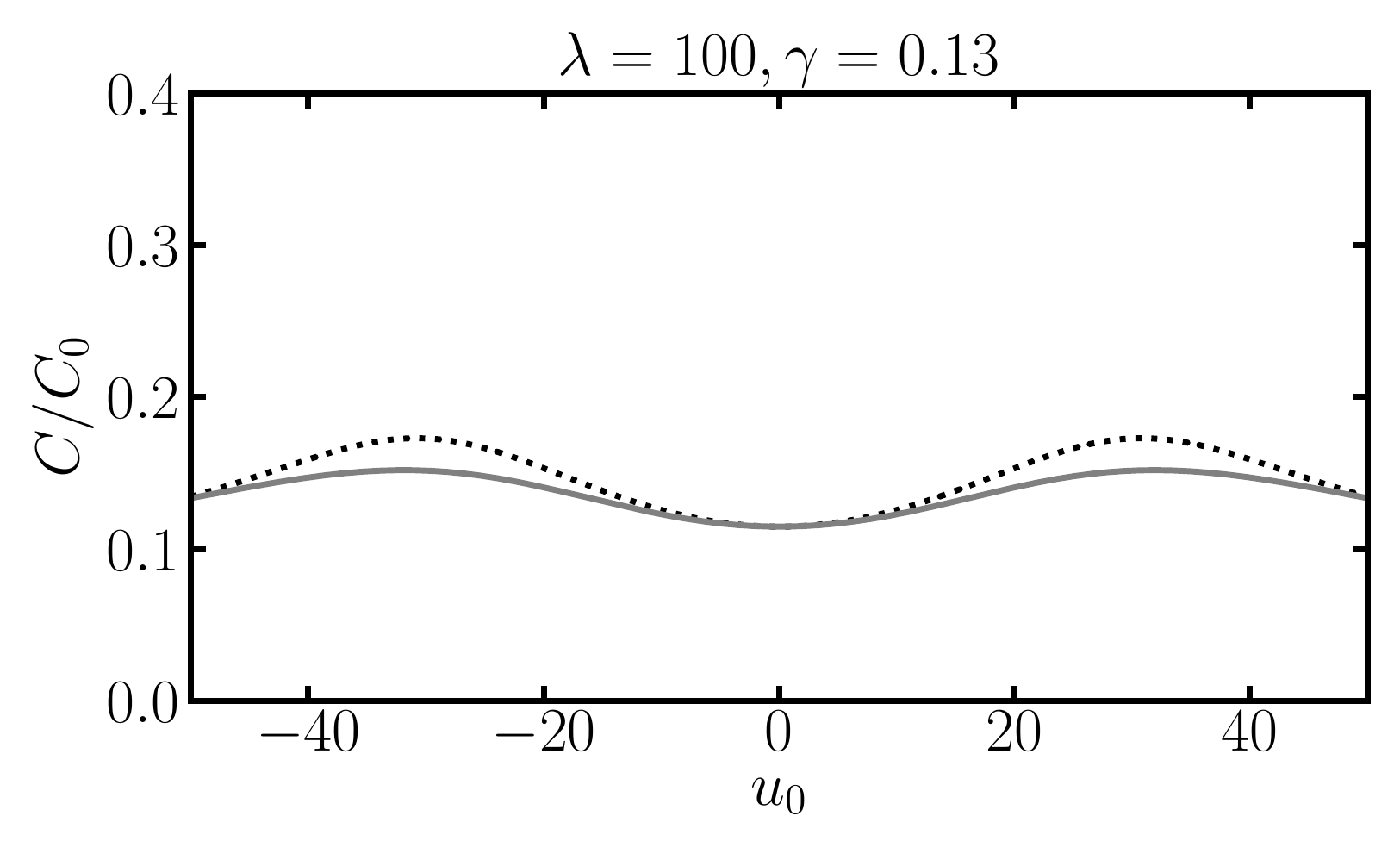}
    \caption{\textbf{The new theory accumulates charge easier than the one just based on free ions}. (left) - charge density, per lattice site, as a function of electrostatic potential, in units of \textcolor{black}{thermal voltage}. (right) - Differential capacitance, in units of Debye capacitance, as a function of electrostatic potential drop across the entire electrical double layer, in units of \textcolor{black}{thermal voltage}. All panels are for the new theory and the free ion theory, as indicated in the titles. Here $\alpha = 0.1$.}
    \label{fig:rho_DC_comp}
\end{figure}

\section{Comparison to free ion theory and experiments}

In Refs.~\citenum{Chen2017,goodwin2017mean,goodwin2017underscreening}, an approximation to ion pairing in the EDL of ILs was developed, which assumed that the ``voids'' were the ``ion pairs''. An equilibrium was established in the bulk to find the proportion of ion pairs. However, Ref.~\citenum{Goodwin2022EDLgel} recently showed that this ion pair equilibrium was not consistently held in the EDL. Therefore, we wish to compare the consistent equations obtained here against that  within this free ion theory (FIT) of Refs.~\citenum{Chen2017,goodwin2017mean,goodwin2017underscreening}. 

In Ref.~\citenum{Chen2017}, the volume fraction of free cations was shown to be 
\begin{equation}
    \bar{\phi}_{\pm} = \dfrac{\gamma e^{\mp\alpha u}/2}{1 - \gamma + \gamma\cosh(\alpha u)}
\end{equation}

\noindent where $\gamma$ is the ``compacity''~\cite{Kornyshev2007}, or in Ref.~\citenum{Chen2017} the fraction of free ions. Here, the free ion fraction is connected to the association probability through $\gamma = 1 - p$. Therefore, the FIT and theory presented here clearly have the same screening length based on the same number of free ions. 

The differential capacitance in Ref.~\citenum{Chen2017} was given by
\begin{equation}
    \dfrac{C}{C_0} = \sqrt{\alpha\gamma}\dfrac{\cosh \left(\alpha u_0/2\right)}{1 + 2\gamma\sinh^2 \left(\alpha u_0/2\right)}\sqrt{\dfrac{2\gamma\sinh^2 \left(\alpha u_0/2\right)}{\ln\left\{1 + 2\gamma\sinh^2 \left(\alpha u_0/2\right)\right\}}},
    \label{eq:cap_free}
\end{equation}

\noindent which is the same as that derived in Refs.~\citenum{Kornyshev2007,kilic2007a}, but with the additional $\alpha$ factor introduced in Ref.~\citenum{goodwin2017mean} that was shown to work well in Ref.~\citenum{BBKGK}. 

In Fig.~\ref{fig:rho_DC_comp}, the new theory is compared against FIT in terms of the charge density, $\rho = \bar{\phi}_+ - \bar{\phi}_-$, and differential capacitance. For $\lambda = 1$ we have $\gamma = 0.73$, which means free ions dominate the IL. There is very little difference between the new theory and FIT, as might be expected when there are few ion pairs. In contrast, for $\lambda = 10$ we have $\gamma = 0.36$. For the new theory, a ``double hump camel'' shape is obtained for the differential capacitance curve, but a ``bell'' shape is obtained for FIT. Both at linear response and at large potential drops across the EDL, the differential capacitance curves and charge densities coincide. Finally, for $\lambda = 100$, which gives $\gamma = 0.13$, both theories predict a ``double hump camel'' differential capacitance curve. Again the new theory has a higher differential capacitance at intermediate voltages.

\textcolor{black}{Overall, both theories coincide at linear response and large potentials, independent of the value of $\lambda$ or $\gamma$. This is expected, since linear response only includes free ions and large potentials is the universal charge conservation law. At intermediate voltages, it is consistently found that the new theory has a larger differential capacitance than the FIT. This is because the charge density in the new theory saturates at $\pm1$ at smaller $u$ than the FIT, creating a larger differential capacitance at these intermediate voltages. This implies that cracking of ion pairs in the EDL causes a more facile accumulation of charge than replacing ion pairs by free ions.} 

\textcolor{black}{This can be further established from our analytical results. In Eq.~\eqref{eq:DC_New}, the ``camel-to-bell'' transition was predicted to occur at a free ion fraction of 1/2. Whereas, in Eq.~\eqref{eq:cap_free} the second order expansion is given by}
\begin{equation}
    \dfrac{C}{C_0} = \sqrt{\alpha\gamma}\left[1 + \dfrac{1 - 3\gamma}{8}(\alpha u_0)^2 \right].
\end{equation}

\noindent \textcolor{black}{which predicts the camel-to-bell transition at $1/3$ of free ions~\cite{Kornyshev2007,kilic2007a,Bazant2009a,Fedorov2014}. Subtracting the coefficient of the quadratic term of the FIT from the new theory, we obtain $3[\sqrt{1 + 2\lambda}(\sqrt{1 + 2\lambda} - 1) - \lambda]/(\lambda\sqrt{1 + 2\lambda}) \geqslant 0$, and therefore, the new theory always has a differential capacitance that is always larger than or equal to the FIT values at small voltages (where this quadratic expansion which holds). For vanishingly small $\lambda$ this coefficient reduces to $0$, as expected.  This coefficient also vanishes at extremely large $\lambda$. For very large $\lambda$ there are very few free ions, and the differential capacitance has a U-shape, similar to that of the Gouy-Chapman theory at small voltages~\cite{Fedorov2014}. The numerical results demonstrates the new theory is always larger than the FIT at intermediate voltages (voltages where the maxima in differential capacitance occur).}


\begin{figure}
    \centering
    \includegraphics[width=0.45\textwidth]{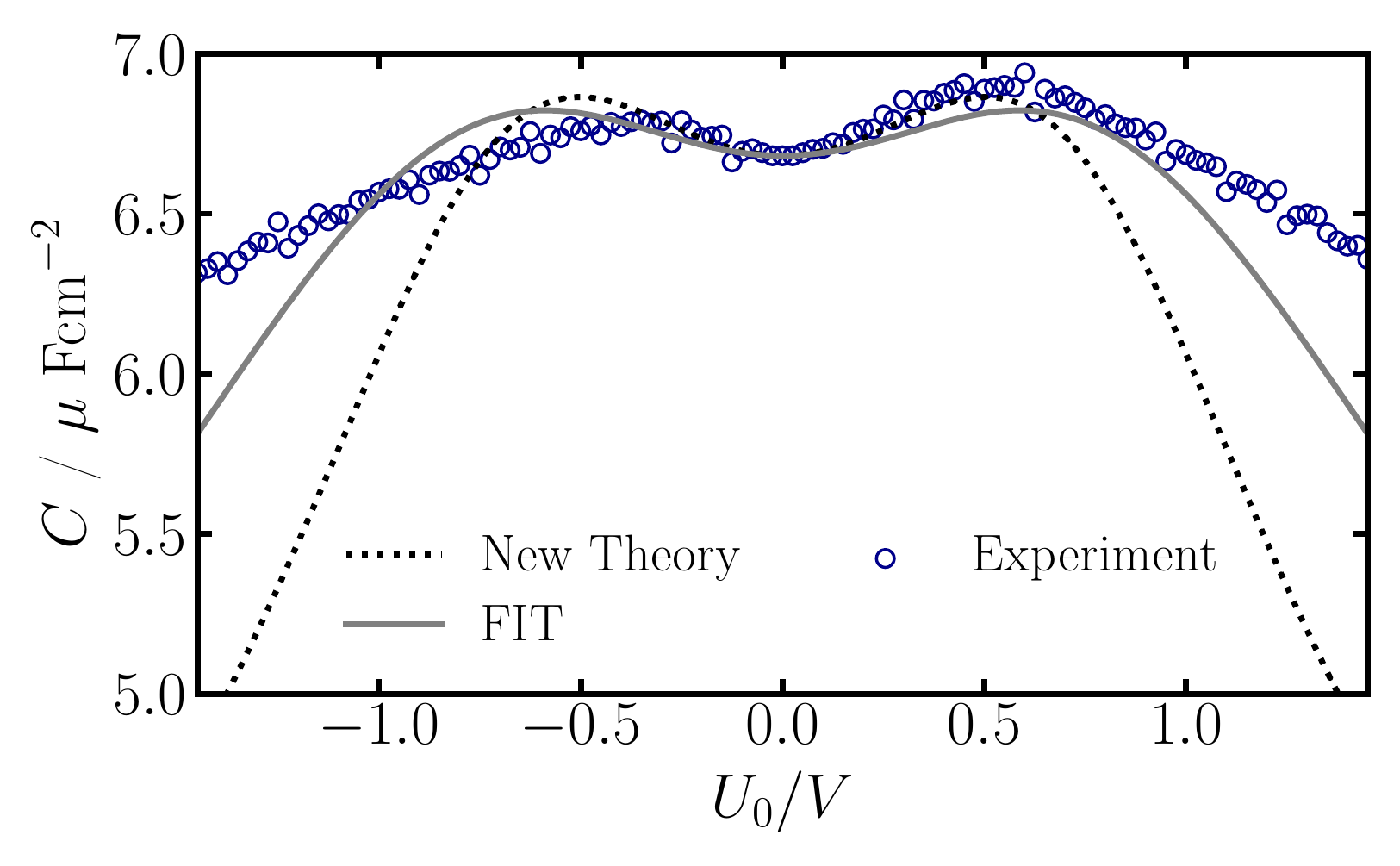}
    \caption{\textbf{The FIT appears to reproduce experimental data better than the new theory, based on a best fit to the data}. The experimental values for [Emim][TFSI] are shown by open circles, and have been reproduced from Ref.~\citenum{Monchai2018}. Both FIT and the new theory have been calculated with $\alpha = 0.0665$ and have had the Debye capacitance rescaled further to match the experimental value, following Jitvisate and Seddon~\cite{Monchai2018}. For FIT, we take $\gamma = 0.278$, as taken in Ref.~\citenum{Monchai2018}. For the new theory, we have taken $\lambda = 8$, to best fit the low-potential curve.}
    \label{fig:EXP_DC}
\end{figure}


\textcolor{black}{Next we turn to comparing these theories against experimental data for the differential capacitance. In Ref.~\citenum{Monchai2018}, Jitvisate and Seddon reported the experimentally obtained differential capacitance curve for 1-Ethyl-3-methylimidazolium bis(trifluoromethylsulfonyl) imide [Emim][TFSI], amongst other ILs, and this IL shall be focused on here as the cation and anion have roughly the same volume. In Fig.~\ref{fig:EXP_DC} the experimental data are reproduced, and compared against the new theory and FIT.}

\textcolor{black}{Based on Ref.~\citenum{goodwin2017mean}, Jitvisate and Seddon~\cite{Monchai2018} fitted their experimental data using the formula represented here as Eq.~\ref{eq:cap_free}. They found $\alpha = 0.0665$ and $\gamma = 0.278$, but noted that the capacitance at the potential of zero charge still had to be reduced. We adopt these best fit values and plot the FIT in Fig.~\ref{fig:EXP_DC}, where the capacitance at the potential of zero charge is further rescaled to exactly match that of the experiments. Overall, the fit is reasonable, with the largest discrepancies occurring at large potentials in the "wings" of the differential capacitance curve.}

\textcolor{black}{In Fig.~\ref{fig:EXP_DC} the new theory is also presented for $\alpha = 0.0665$ and $\lambda = 8$, which is our best fit to the fata, again where the Debye capacitance has been further reduced to match that of the experiments. The new theory reasonably captures the small voltage differential capacitance, but the wings of the differential capacitance reduces substantially below that of the experiment for voltages larger than $\pm1$~V. }

\textcolor{black}{Overall, the FIT theory does a better job at reproducing the experimental data for [Emim][TFSI] than the new theory, based on a best fit to the data. This is because the new theory has a larger propensity to form a ``double hump camel'' shape than the FIT. Moreover, as we have already mentioned, Feng \textit{et al.}~\cite{feng2019free} found that the fraction of free ions for [Emim][TFSI] should be approximately 0.15. The fraction of free ions predicted by the new theory is 0.39, which is even larger than that predicted by the FIT of 0.28. Therefore, the new theory is also in worse agreement in terms of the fraction of free ions than FIT. Note that the reason why the new theory has its wings of differential capacitance lower than that of the FIT is because there is a larger free ion fraction which means the crowding regime is reached at smaller values of the potential drop across the EDL (and also because a smaller $C_0$ is required to obtain the experimental capacitance at the potential of zero charge).}

\textcolor{black}{Both of these deficiencies of the new theory can be attributed to limiting the functionality to ion pairs. In Ref.~\citenum{mceldrew2020correlated}, the functionality of both the cation and anion were found to be 4. In Ref.~\citenum{Goodwin2022EDLgel}, this functionality was used to calculate the differential capacitance, and better agreement was found with experiments in terms of the differential capacitance (in comparison to FIT) when using a free ion fraction close to that of Feng \textit{et al.}~\cite{feng2019free}. This demonstrates \textit{the importance of knowing the functionality of ions accurately}. Moreover, this is also important for the interpretation of experiments, as Gebbie \textit{et al.}~\cite{Gebbie2013,Gebbie2015} suggested that over 99.99\% of the ions were bound up in neutral ion pairs, but McEldrew \textit{et al.}~\cite{mceldrew2020correlated} showed that these ions have functionalities of 4-5. The differential capacitance and free ion fractions for these functionalities (1 vs 4-5) are very different, and can result in qualitatively different predictions.}

\textcolor{black}{Despite the new theory under-performing against FIT for [Emim][TFSI], this does not mean that the new theory is irrelevant. The presented version should accurately work when the ions of the IL can only form ion pairs, and should provide a starting point to further investigate correlation and non-equilibrium effects.}

\section{Ion pair orientation and the Wien effect}

Zhang \textit{et al.}~\cite{yufan2020} developed a FIT which accounted for ion pair orientation in the EDL, based on fluctuating Langevin dipoles. While the lifetime of an association has been found to be of the order of 1-10~ps~\cite{Zhang2015,mceldrew2020correlated}, which means the association lifetimes are presumably shorter than the timescale of rotation for an ion pair. Nonetheless, it is interesting to investigate how one would introduce ion pairs as fluctuating Langevin dipoles within the theory presented here. This could help develop more sophisticated approaches which include the dielectric response of the IL from clusters beyond ion pairs. 

The free energy contribution for ion pairs acting as fluctuating Langevin dipoles is given by
\begin{equation}
    \beta \mathcal{F} = N_{11}\ln\left\{\dfrac{\sinh \left( \tilde{p}|\tilde{\nabla} u|\right)}{\tilde{p}|\tilde{\nabla} u|}\right\},
\end{equation}

\noindent where $\tilde{p}$ is the dimensionless dipole moment of the ion pair. Note this also modifies the Poisson equation, such that the dielectric function depends on the electric field. Establishing chemical equilibrium between free ions and ion pairs within the EDL, we find that the association constant acquires an additional contribution which depends on the electric field
\begin{equation}
    \bar{\lambda} = \lambda\dfrac{\sinh \left( \tilde{p}|\nabla u|\right)}{\tilde{p}|\nabla u|}.
\end{equation}

\noindent This additional electric field dependence does not alter linear response, but for larger fields, it causes the association constant to increase in the EDL, and therefore, the IL could become more associated in the EDL. This will be in competition with the changing volume fractions of cations and anions which tends to reduce the total number of associations. The unequal numbers of cations and anions should dominate over the increase in $\bar{\lambda}$, but numerical calculations will be required to understand if this is the case. 

It is expected that this does not occur to a substantial extent in ILs, however. This is because the ion pairs have a short lifetime~\cite{Zhang2015,mceldrew2020correlated}, and so will not behave as ideal fluctuating Langevin dipoles. Moreover, even if they did, the orientation of the ion pair along the electric field would cause the ion pair to be stretched, reducing the binding energy of the ions, and eventually resulting in an association constant which tends to zero. For functionalities larger than 1, where Cayley tree clusters form, introducing the orientation of these larger clusters becomes difficult, and could result in a break-down of the assumed cluster distribution. 

\textcolor{black}{The electric field dependence of the (non-equilibrium) association/dissociation constant of ion pairs is actually a well-known phenomenon~\cite{Onsager1934,Kaiser2013}. In weakly dissociating acids and electrolytes, it is referred to as the second Wien effect~\cite{Onsager1934,Kaiser2013} (the Wien effect is the increase of the electrical conductivity in large electrostatic fields from changes to the ionic atmosphere and dissociation of ion pairs~\cite{Shoon1957,Kaiser2013}). It was shown by Onsager that the (non-equilibrium) dissociation constant of Bjerrum pairs depends on the electric field from the dynamical exchange of ions between the free state and Bjerrum paired state, since the association of an ion pair is independent of field but the dissociation of an ion pair depends on the field~\cite{Onsager1934,Kaiser2013}. Therefore, the dissociation constant increases with increasing electric field, i.e., the liberation of free ions with increasing electrostatic field~\cite{Onsager1934,Kaiser2013}. Such effect should also occur in ILs~\cite{Kumar2015,Patro2016,Roling2017}, as there is a dynamical exchange between the two states. Therefore, it is expected that the equilibrium constant has a dependence on the electric field which drives ions to the free state.}

\textcolor{black}{In the context of ILs, there has been some investigation into the Wien effect~\cite{Patro2016}. The Walden plot is typically used to quantify if an electrolyte is strongly or weakly dissociating~\cite{MacFarlane2009,Patro2016,Roling2017}. Some ILs have conductivities close to the ideal line, and are considered as strong electrolytes, but there are quite a number of examples which deviate from ideality of the Nernst-Einstein and are considered as weakly dissociated~\cite{MacFarlane2009,Patro2016,Roling2017}. This is also known as the ionicity in ILs~\cite{MacFarlane2009}.}




\section{Discussion}


It was shown by Downing \textit{et al.}~\cite{Downing2018}, using a quasi-chemical approximation (QCA), that correlations between pairs of ions in an incompressible IL causes the differential capacitance curve to be more ``double hump camel''-like. This QCA treats short-ranged correlations between pairs of ions through the formation of ion pairs. Note that cation-cation and anion-anion pairs were also permitted. This higher tendency for ``double hump camel'' shaped differential capacitance curves is consistent with the theory presented here, despite their theory intrinsically linking the regular solution correlations with associations, whilst our approach treats short-range associations orthogonal to short-range regular solution repulsion. It was also noted by Downing \textit{et al.}~\cite{Downing2018} that the QCA can be extended beyond ion pairs, and to more ordered aggregates~\cite{Bossa2015}, which could be a way to model electrolytes which do not form Cayley tree clusters, such as NaCl~\cite{choi2018graph}. Also see Ref.~\citenum{marcus2006ion} for a detailed review of other mehods for modelling ion pair formation and associations.



It was shown by McEldrew \textit{et al.}~\cite{mceldrew2020correlated,mceldrew2021salt} that some typical ILs form Cayley tree clusters with a functionality between 3 and 5. Moreover, the functionality of the cation is typically different to the anion ($f_- \neq f_+$), in addition to their volumes being different ($v_- \neq v_+$). It was shown that the presented theory does worse than the more general theory which accounts for these larger functionalities, which highlights the importance of accurately knowing the functionality. Despite this, the presented theory is still a conceptually useful tool, and ILs which only form ion pairs could still be found. Moreover, the presented theory could be developed further to apply to other paired systems, such as weakly dissociating acids.


In Ref.~\citenum{Goodwin2022EDLgel} some issues of the presented approach were highlighted. For example, the fact that a local density approach is utilised, and the screening length is typically shorter than the size of the clusters. Here, provided there are significant numbers of ion pairs, the screening length is larger than the characteristic size of an ion pair. This means a local density approximation should capture the mean-field volume fractions of species in the EDL. However, the internal structure of the ion pair is not captured here. It was shown by Avni \textit{et al.}~\cite{avni2020charge}, that ion pair formation (and other small, ordered aggregates) gives rise to overscreening equations. This indicated that ion pairing is conceptually similar to overscreening. This was also established by Ma \textit{et al.}~\cite{Ma2015}, where it was noted the overscreening structure at small electrode potentials hardly changes, even if 99.99\% of ions are free. Further development of the theory to account for the internal structure of ion pairs and electrostatic correlations could further strengthen the link between overscreening and ion pair formation. 




\section{Conclusion}

In summary, we have investigated the role of ion pairing in the electrical double layer (EDL) of ionic liquids (IL), which consistently accounts for the chemical equilibria in the EDL, bulk, and the equilibrium between the two. This conceptually simple case permitted analytical solutions to be obtained for the charge density, surface charge density and differential capacitance, which shed light on the equilibrium between the free energy of an association and the electrostatic potential energy of an ion. The effect of ion pairs acting as fluctuating Langevin dipoles was also touched upon, and we found that this would tend to increase the number of associations in the EDL. \textcolor{black}{Conceptually, this is opposite to the field-effect dissociation of weak acids, which is also known as the second Wien effect.}

\textcolor{black}{We also compared the newly derived equation for differential capacitance against that of previous free ion approaches. Overall, we found that the free ion approaches can actually fit experimental data better than the new equation for differential capacitance. However, both the new theory and free ion approaches do significantly worse than the more general theory (Ref.~\citenum{Goodwin2022EDLgel}) which has a functionality of the ions determined from molecular dynamics simulations~\cite{mceldrew2020correlated}. If the derived differential capacitance equation is used to fit experimental data, it will tend to overestimate the number of free ions and the wings in differential capacitance will decrease more than expected as a result. Therefore, \textit{it is key to know the functionality of ions accurately if one is to predict their properties}.} 

\textcolor{black}{The value of the presented analysis is in the analytical equations which were derived. From these equations, we learnt the following points:}

\begin{enumerate}
    \item \textcolor{black}{The charge density [or volume fraction of cations - Eq.~\eqref{eq:EDL_phip}] show a clear competition between the free energy of an association and the electrostatic potential. When $-\beta\Delta f_{+-} > 2\alpha|u|$ associations dominate, but when $2\alpha|u| > -\beta\Delta f_{+-}$ there is a substantial field-effect cracking of ion pairs into free ions.}
    \item \textcolor{black}{The obtained differential capacitance has a larger propensity to form a ``double hump'' differential capacitance curve. The ``camel-to-bell'' transition was predicted to occur when the free ion fraction equals 1/2, as shown by Eq.~\eqref{eq:DC_New}.}
    \item \textcolor{black}{In comparison to free ion approaches, the differential capacitance of the new theory is always larger at intermediate voltages, as revealed by the Taylor expansions of the differential capacitance equations.}
\end{enumerate}

\textcolor{black}{All in all, this work is, to our knowledge, the first analytical theory of the electric-field-induced dissociation of ions pairs in EDL. It has resulted in a simple formula for the dissociation law and the EDL capacitance that account for such dissociation. The new expression for capacitance contains the key factors entering the theory: the propensity to pair, short range correlations parameter and crowding. As it is, it has been applied for the case of ILs, it comes out as not sufficient, because ions there tend to form clusters of all ranks, not just `ion pairs'. The presented theory serves as a starting point for further development to include non-local electrostatic correlations, for applications in non-equilibrium effects, and to conventional electrolytes through including solvent in the theory. One application for the last point could be to salts dissolved in low dielectric constant solvents, where ion pairing have been suggested to occur~\cite{IBL}. }


\section{Acknowledgements}


We acknowledge stimulating discussions with J. Pedro de Souza, Micheal McEldrew and Martin Bazant. ZG was supported through a studentship in the Centre for Doctoral Training on Theory and Simulation of Materials at Imperial College London funded by the EPSRC (EP/L015579/1) and from the Thomas Young Centre under grant number TYC-101. AK would like to acknowledge the research grant by the Leverhulme Trust (RPG-2016- 223). 

\bibliography{main.bib}

\begin{thebibliography}{108}%
\makeatletter
\providecommand \@ifxundefined [1]{%
 \@ifx{#1\undefined}
}%
\providecommand \@ifnum [1]{%
 \ifnum #1\expandafter \@firstoftwo
 \else \expandafter \@secondoftwo
 \fi
}%
\providecommand \@ifx [1]{%
 \ifx #1\expandafter \@firstoftwo
 \else \expandafter \@secondoftwo
 \fi
}%
\providecommand \natexlab [1]{#1}%
\providecommand \enquote  [1]{``#1''}%
\providecommand \bibnamefont  [1]{#1}%
\providecommand \bibfnamefont [1]{#1}%
\providecommand \citenamefont [1]{#1}%
\providecommand \href@noop [0]{\@secondoftwo}%
\providecommand \href [0]{\begingroup \@sanitize@url \@href}%
\providecommand \@href[1]{\@@startlink{#1}\@@href}%
\providecommand \@@href[1]{\endgroup#1\@@endlink}%
\providecommand \@sanitize@url [0]{\catcode `\\12\catcode `\$12\catcode
  `\&12\catcode `\#12\catcode `\^12\catcode `\_12\catcode `\%12\relax}%
\providecommand \@@startlink[1]{}%
\providecommand \@@endlink[0]{}%
\providecommand \url  [0]{\begingroup\@sanitize@url \@url }%
\providecommand \@url [1]{\endgroup\@href {#1}{\urlprefix }}%
\providecommand \urlprefix  [0]{URL }%
\providecommand \Eprint [0]{\href }%
\providecommand \doibase [0]{http://dx.doi.org/}%
\providecommand \selectlanguage [0]{\@gobble}%
\providecommand \bibinfo  [0]{\@secondoftwo}%
\providecommand \bibfield  [0]{\@secondoftwo}%
\providecommand \translation [1]{[#1]}%
\providecommand \BibitemOpen [0]{}%
\providecommand \bibitemStop [0]{}%
\providecommand \bibitemNoStop [0]{.\EOS\space}%
\providecommand \EOS [0]{\spacefactor3000\relax}%
\providecommand \BibitemShut  [1]{\csname bibitem#1\endcsname}%
\let\auto@bib@innerbib\@empty
\bibitem [{\citenamefont {Welton}(1999)}]{Welton1999}%
  \BibitemOpen
  \bibfield  {author} {\bibinfo {author} {\bibfnamefont {T.}~\bibnamefont
  {Welton}},\ }\href@noop {} {\bibfield  {journal} {\bibinfo  {journal} {Chem.
  Rev.}\ }\textbf {\bibinfo {volume} {99}},\ \bibinfo {pages} {2071} (\bibinfo
  {year} {1999})}\BibitemShut {NoStop}%
\bibitem [{\citenamefont {Weing\"{a}rtner}(2008)}]{Hermann2008}%
  \BibitemOpen
  \bibfield  {author} {\bibinfo {author} {\bibfnamefont {H.}~\bibnamefont
  {Weing\"{a}rtner}},\ }\href@noop {} {\bibfield  {journal} {\bibinfo
  {journal} {Angew. Chem. Int. Ed.}\ }\textbf {\bibinfo {volume} {47}},\
  \bibinfo {pages} {654} (\bibinfo {year} {2008})}\BibitemShut {NoStop}%
\bibitem [{\citenamefont {Hallett}\ and\ \citenamefont
  {Welton}(2011)}]{Hallett2011}%
  \BibitemOpen
  \bibfield  {author} {\bibinfo {author} {\bibfnamefont {J.~P.}\ \bibnamefont
  {Hallett}}\ and\ \bibinfo {author} {\bibfnamefont {T.}~\bibnamefont
  {Welton}},\ }\href@noop {} {\bibfield  {journal} {\bibinfo  {journal} {Chem.
  Rev.}\ }\textbf {\bibinfo {volume} {111}},\ \bibinfo {pages} {3508} (\bibinfo
  {year} {2011})}\BibitemShut {NoStop}%
\bibitem [{\citenamefont {Kondrat}\ and\ \citenamefont
  {Kornyshev}(2016)}]{Kondrat2016}%
  \BibitemOpen
  \bibfield  {author} {\bibinfo {author} {\bibfnamefont {S.}~\bibnamefont
  {Kondrat}}\ and\ \bibinfo {author} {\bibfnamefont {A.~A.}\ \bibnamefont
  {Kornyshev}},\ }\href@noop {} {\bibfield  {journal} {\bibinfo  {journal}
  {Nanoscale Horiz.}\ }\textbf {\bibinfo {volume} {1}},\ \bibinfo {pages} {45}
  (\bibinfo {year} {2016})}\BibitemShut {NoStop}%
\bibitem [{\citenamefont {Son}\ and\ \citenamefont {Wang}(2020)}]{son2020ion}%
  \BibitemOpen
  \bibfield  {author} {\bibinfo {author} {\bibfnamefont {C.~Y.}\ \bibnamefont
  {Son}}\ and\ \bibinfo {author} {\bibfnamefont {Z.-G.}\ \bibnamefont {Wang}},\
  }\href@noop {} {\bibfield  {journal} {\bibinfo  {journal} {J. Chem. Phys.}\
  }\textbf {\bibinfo {volume} {153}},\ \bibinfo {pages} {100903} (\bibinfo
  {year} {2020})}\BibitemShut {NoStop}%
\bibitem [{\citenamefont {Fedorov}\ and\ \citenamefont
  {Kornyshev}(2014)}]{Fedorov2014}%
  \BibitemOpen
  \bibfield  {author} {\bibinfo {author} {\bibfnamefont {M.~V.}\ \bibnamefont
  {Fedorov}}\ and\ \bibinfo {author} {\bibfnamefont {A.~A.}\ \bibnamefont
  {Kornyshev}},\ }\href {\doibase 10.1021/cr400374x} {\bibfield  {journal}
  {\bibinfo  {journal} {Chem. Rev.}\ }\textbf {\bibinfo {volume} {114}},\
  \bibinfo {pages} {2978} (\bibinfo {year} {2014})}\BibitemShut {NoStop}%
\bibitem [{\citenamefont {Ivani\v{s}t\v{s}ev}\ \emph
  {et~al.}(2015)\citenamefont {Ivani\v{s}t\v{s}ev}, \citenamefont {Kirchner},
  \citenamefont {Kirchner},\ and\ \citenamefont {Fedorov}}]{Fedorov2015Re}%
  \BibitemOpen
  \bibfield  {author} {\bibinfo {author} {\bibfnamefont {V.}~\bibnamefont
  {Ivani\v{s}t\v{s}ev}}, \bibinfo {author} {\bibfnamefont {K.}~\bibnamefont
  {Kirchner}}, \bibinfo {author} {\bibfnamefont {T.}~\bibnamefont {Kirchner}},
  \ and\ \bibinfo {author} {\bibfnamefont {M.~V.}\ \bibnamefont {Fedorov}},\
  }\href@noop {} {\bibfield  {journal} {\bibinfo  {journal} {J. Phys.: Condens.
  Matter}\ }\textbf {\bibinfo {volume} {27}},\ \bibinfo {pages} {102101}
  (\bibinfo {year} {2015})}\BibitemShut {NoStop}%
\bibitem [{\citenamefont {Trulsson}\ \emph {et~al.}(2010)\citenamefont
  {Trulsson}, \citenamefont {Algotsson}, \citenamefont {Forsman},\ and\
  \citenamefont {Woodward}}]{Trulsson2010}%
  \BibitemOpen
  \bibfield  {author} {\bibinfo {author} {\bibfnamefont {M.}~\bibnamefont
  {Trulsson}}, \bibinfo {author} {\bibfnamefont {J.}~\bibnamefont {Algotsson}},
  \bibinfo {author} {\bibfnamefont {J.}~\bibnamefont {Forsman}}, \ and\
  \bibinfo {author} {\bibfnamefont {C.~E.}\ \bibnamefont {Woodward}},\
  }\href@noop {} {\bibfield  {journal} {\bibinfo  {journal} {J Phys. Chem.
  Lett.}\ }\textbf {\bibinfo {volume} {1}},\ \bibinfo {pages} {1191} (\bibinfo
  {year} {2010})}\BibitemShut {NoStop}%
\bibitem [{\citenamefont {Sha}\ \emph {et~al.}(2014)\citenamefont {Sha},
  \citenamefont {Dou}, \citenamefont {Luo}, \citenamefont {Zhu},\ and\
  \citenamefont {Wu}}]{Sha2014}%
  \BibitemOpen
  \bibfield  {author} {\bibinfo {author} {\bibfnamefont {M.}~\bibnamefont
  {Sha}}, \bibinfo {author} {\bibfnamefont {Q.}~\bibnamefont {Dou}}, \bibinfo
  {author} {\bibfnamefont {F.}~\bibnamefont {Luo}}, \bibinfo {author}
  {\bibfnamefont {G.}~\bibnamefont {Zhu}}, \ and\ \bibinfo {author}
  {\bibfnamefont {G.}~\bibnamefont {Wu}},\ }\href@noop {} {\bibfield  {journal}
  {\bibinfo  {journal} {ACS Appl. Mater. Interfaces}\ }\textbf {\bibinfo
  {volume} {6}},\ \bibinfo {pages} {12556} (\bibinfo {year}
  {2014})}\BibitemShut {NoStop}%
\bibitem [{\citenamefont {Vatamanu}\ \emph {et~al.}(2012)\citenamefont
  {Vatamanu}, \citenamefont {Borodin}, \citenamefont {Bedrov},\ and\
  \citenamefont {Smith}}]{Vatamanu2012}%
  \BibitemOpen
  \bibfield  {author} {\bibinfo {author} {\bibfnamefont {J.}~\bibnamefont
  {Vatamanu}}, \bibinfo {author} {\bibfnamefont {O.}~\bibnamefont {Borodin}},
  \bibinfo {author} {\bibfnamefont {D.}~\bibnamefont {Bedrov}}, \ and\ \bibinfo
  {author} {\bibfnamefont {G.~D.}\ \bibnamefont {Smith}},\ }\href {\doibase
  10.1021/jp301399b} {\bibfield  {journal} {\bibinfo  {journal} {The Journal of
  Physical Chemistry C}\ }\textbf {\bibinfo {volume} {116}},\ \bibinfo {pages}
  {7940} (\bibinfo {year} {2012})}\BibitemShut {NoStop}%
\bibitem [{\citenamefont {Merlet}\ \emph {et~al.}(2014)\citenamefont {Merlet},
  \citenamefont {Limmer}, \citenamefont {Salanne}, \citenamefont {van Roij},
  \citenamefont {Madden}, \citenamefont {Chandler},\ and\ \citenamefont
  {Rotenberg}}]{Merlet2014}%
  \BibitemOpen
  \bibfield  {author} {\bibinfo {author} {\bibfnamefont {C.}~\bibnamefont
  {Merlet}}, \bibinfo {author} {\bibfnamefont {D.~T.}\ \bibnamefont {Limmer}},
  \bibinfo {author} {\bibfnamefont {M.}~\bibnamefont {Salanne}}, \bibinfo
  {author} {\bibfnamefont {R.}~\bibnamefont {van Roij}}, \bibinfo {author}
  {\bibfnamefont {P.~A.}\ \bibnamefont {Madden}}, \bibinfo {author}
  {\bibfnamefont {D.}~\bibnamefont {Chandler}}, \ and\ \bibinfo {author}
  {\bibfnamefont {B.}~\bibnamefont {Rotenberg}},\ }\href {\doibase
  10.1021/jp503224w} {\bibfield  {journal} {\bibinfo  {journal} {The Journal of
  Physical Chemistry C}\ }\textbf {\bibinfo {volume} {118}},\ \bibinfo {pages}
  {18291} (\bibinfo {year} {2014})}\BibitemShut {NoStop}%
\bibitem [{\citenamefont {Merlet}\ \emph {et~al.}(2013)\citenamefont {Merlet},
  \citenamefont {Rotenberg}, \citenamefont {Madden},\ and\ \citenamefont
  {Salanne}}]{Merlet2013}%
  \BibitemOpen
  \bibfield  {author} {\bibinfo {author} {\bibfnamefont {C.}~\bibnamefont
  {Merlet}}, \bibinfo {author} {\bibfnamefont {B.}~\bibnamefont {Rotenberg}},
  \bibinfo {author} {\bibfnamefont {P.~A.}\ \bibnamefont {Madden}}, \ and\
  \bibinfo {author} {\bibfnamefont {M.}~\bibnamefont {Salanne}},\ }\href
  {\doibase 10.1039/c3cp52088a} {\bibfield  {journal} {\bibinfo  {journal}
  {Physical Chemistry Chemical Physics}\ }\textbf {\bibinfo {volume} {15}},\
  \bibinfo {pages} {15781} (\bibinfo {year} {2013})}\BibitemShut {NoStop}%
\bibitem [{\citenamefont {Merlet}\ \emph {et~al.}(2011)\citenamefont {Merlet},
  \citenamefont {Salanne}, \citenamefont {Rotenberg},\ and\ \citenamefont
  {Madden}}]{Merlet2011}%
  \BibitemOpen
  \bibfield  {author} {\bibinfo {author} {\bibfnamefont {C.}~\bibnamefont
  {Merlet}}, \bibinfo {author} {\bibfnamefont {M.}~\bibnamefont {Salanne}},
  \bibinfo {author} {\bibfnamefont {B.}~\bibnamefont {Rotenberg}}, \ and\
  \bibinfo {author} {\bibfnamefont {P.~A.}\ \bibnamefont {Madden}},\ }\href
  {\doibase 10.1021/jp205461g} {\bibfield  {journal} {\bibinfo  {journal} {The
  Journal of Physical Chemistry C}\ }\textbf {\bibinfo {volume} {115}},\
  \bibinfo {pages} {16613} (\bibinfo {year} {2011})}\BibitemShut {NoStop}%
\bibitem [{\citenamefont {Bhuiyan}\ \emph {et~al.}(2012)\citenamefont
  {Bhuiyan}, \citenamefont {Lamperski}, \citenamefont {Wu},\ and\ \citenamefont
  {Henderson}}]{bhuiyan2012monte}%
  \BibitemOpen
  \bibfield  {author} {\bibinfo {author} {\bibfnamefont {L.~B.}\ \bibnamefont
  {Bhuiyan}}, \bibinfo {author} {\bibfnamefont {S.}~\bibnamefont {Lamperski}},
  \bibinfo {author} {\bibfnamefont {J.}~\bibnamefont {Wu}}, \ and\ \bibinfo
  {author} {\bibfnamefont {D.}~\bibnamefont {Henderson}},\ }\href@noop {}
  {\bibfield  {journal} {\bibinfo  {journal} {J. Phys. Chem. B}\ }\textbf
  {\bibinfo {volume} {116}},\ \bibinfo {pages} {10364} (\bibinfo {year}
  {2012})}\BibitemShut {NoStop}%
\bibitem [{\citenamefont {Lamperski}\ \emph {et~al.}(2014)\citenamefont
  {Lamperski}, \citenamefont {Sosnowska}, \citenamefont {Bhuiyan},\ and\
  \citenamefont {Henderson}}]{Lamperski2014}%
  \BibitemOpen
  \bibfield  {author} {\bibinfo {author} {\bibfnamefont {S.}~\bibnamefont
  {Lamperski}}, \bibinfo {author} {\bibfnamefont {J.}~\bibnamefont
  {Sosnowska}}, \bibinfo {author} {\bibfnamefont {L.~B.}\ \bibnamefont
  {Bhuiyan}}, \ and\ \bibinfo {author} {\bibfnamefont {D.}~\bibnamefont
  {Henderson}},\ }\href@noop {} {\bibfield  {journal} {\bibinfo  {journal} {J.
  Chem. Phys.}\ }\textbf {\bibinfo {volume} {140}},\ \bibinfo {pages} {014704}
  (\bibinfo {year} {2014})}\BibitemShut {NoStop}%
\bibitem [{\citenamefont {Forsman}\ \emph {et~al.}(2011)\citenamefont
  {Forsman}, \citenamefont {Woodward},\ and\ \citenamefont
  {Trulsson}}]{Forsman2011}%
  \BibitemOpen
  \bibfield  {author} {\bibinfo {author} {\bibfnamefont {J.}~\bibnamefont
  {Forsman}}, \bibinfo {author} {\bibfnamefont {C.~E.}\ \bibnamefont
  {Woodward}}, \ and\ \bibinfo {author} {\bibfnamefont {M.}~\bibnamefont
  {Trulsson}},\ }\href@noop {} {\bibfield  {journal} {\bibinfo  {journal} {J.
  Phys. Chem. B}\ }\textbf {\bibinfo {volume} {115}},\ \bibinfo {pages} {4606}
  (\bibinfo {year} {2011})}\BibitemShut {NoStop}%
\bibitem [{\citenamefont {Gavish}\ and\ \citenamefont
  {Yochelis}(2016)}]{gavish2016}%
  \BibitemOpen
  \bibfield  {author} {\bibinfo {author} {\bibfnamefont {N.}~\bibnamefont
  {Gavish}}\ and\ \bibinfo {author} {\bibfnamefont {A.}~\bibnamefont
  {Yochelis}},\ }\href@noop {} {\bibfield  {journal} {\bibinfo  {journal}
  {Journal Phys. Chem. Lett.}\ }\textbf {\bibinfo {volume} {7}},\ \bibinfo
  {pages} {1121} (\bibinfo {year} {2016})}\BibitemShut {NoStop}%
\bibitem [{\citenamefont {Han}\ \emph {et~al.}(2014)\citenamefont {Han},
  \citenamefont {Huang},\ and\ \citenamefont {Yan}}]{Han2014}%
  \BibitemOpen
  \bibfield  {author} {\bibinfo {author} {\bibfnamefont {Y.}~\bibnamefont
  {Han}}, \bibinfo {author} {\bibfnamefont {S.}~\bibnamefont {Huang}}, \ and\
  \bibinfo {author} {\bibfnamefont {T.}~\bibnamefont {Yan}},\ }\href {\doibase
  10.1088/0953-8984/26/28/284103} {\bibfield  {journal} {\bibinfo  {journal}
  {J. Phys.: Condens. Matter}\ }\textbf {\bibinfo {volume} {26}},\ \bibinfo
  {pages} {284103} (\bibinfo {year} {2014})}\BibitemShut {NoStop}%
\bibitem [{\citenamefont {Maggs}\ and\ \citenamefont
  {Podgornik}(2016)}]{Maggs2016}%
  \BibitemOpen
  \bibfield  {author} {\bibinfo {author} {\bibfnamefont {A.~C.}\ \bibnamefont
  {Maggs}}\ and\ \bibinfo {author} {\bibfnamefont {R.}~\bibnamefont
  {Podgornik}},\ }\href {\doibase 10.1039/C5SM01757B} {\bibfield  {journal}
  {\bibinfo  {journal} {Soft Matter}\ }\textbf {\bibinfo {volume} {12}},\
  \bibinfo {pages} {1219} (\bibinfo {year} {2016})}\BibitemShut {NoStop}%
\bibitem [{\citenamefont {Girotto}\ \emph {et~al.}(2017)\citenamefont
  {Girotto}, \citenamefont {Colla}, \citenamefont {dos Santos},\ and\
  \citenamefont {Levin}}]{Girotto2017}%
  \BibitemOpen
  \bibfield  {author} {\bibinfo {author} {\bibfnamefont {M.}~\bibnamefont
  {Girotto}}, \bibinfo {author} {\bibfnamefont {T.}~\bibnamefont {Colla}},
  \bibinfo {author} {\bibfnamefont {A.~P.}\ \bibnamefont {dos Santos}}, \ and\
  \bibinfo {author} {\bibfnamefont {Y.}~\bibnamefont {Levin}},\ }\href@noop {}
  {\bibfield  {journal} {\bibinfo  {journal} {J. Phys. Chem. B}\ }\textbf
  {\bibinfo {volume} {121}},\ \bibinfo {pages} {6408} (\bibinfo {year}
  {2017})}\BibitemShut {NoStop}%
\bibitem [{\citenamefont {Limmer}(2015)}]{Limmer2015}%
  \BibitemOpen
  \bibfield  {author} {\bibinfo {author} {\bibfnamefont {D.~T.}\ \bibnamefont
  {Limmer}},\ }\href@noop {} {\bibfield  {journal} {\bibinfo  {journal} {Phys.
  Rev. Lett.}\ }\textbf {\bibinfo {volume} {115}},\ \bibinfo {pages} {256102}
  (\bibinfo {year} {2015})}\BibitemShut {NoStop}%
\bibitem [{\citenamefont {Lauw}\ \emph {et~al.}(2009)\citenamefont {Lauw},
  \citenamefont {Horne}, \citenamefont {Rodopoulos},\ and\ \citenamefont
  {Leermakers}}]{Lauw2009}%
  \BibitemOpen
  \bibfield  {author} {\bibinfo {author} {\bibfnamefont {Y.}~\bibnamefont
  {Lauw}}, \bibinfo {author} {\bibfnamefont {M.~D.}\ \bibnamefont {Horne}},
  \bibinfo {author} {\bibfnamefont {T.}~\bibnamefont {Rodopoulos}}, \ and\
  \bibinfo {author} {\bibfnamefont {F.~A.~M.}\ \bibnamefont {Leermakers}},\
  }\href@noop {} {\bibfield  {journal} {\bibinfo  {journal} {Phys. Rev. Lett.}\
  }\textbf {\bibinfo {volume} {103}},\ \bibinfo {pages} {117801} (\bibinfo
  {year} {2009})}\BibitemShut {NoStop}%
\bibitem [{\citenamefont {Levin}(2002)}]{Levin2002}%
  \BibitemOpen
  \bibfield  {author} {\bibinfo {author} {\bibfnamefont {Y.}~\bibnamefont
  {Levin}},\ }\href@noop {} {\bibfield  {journal} {\bibinfo  {journal} {Rep.
  Prog. Phys.}\ }\textbf {\bibinfo {volume} {65}},\ \bibinfo {pages}
  {1577–1632} (\bibinfo {year} {2002})}\BibitemShut {NoStop}%
\bibitem [{\citenamefont {Goodwin}\ \emph {et~al.}(2021)\citenamefont
  {Goodwin}, \citenamefont {de~Souza}, \citenamefont {Bazant},\ and\
  \citenamefont {Kornyshev}}]{goodwin2021review}%
  \BibitemOpen
  \bibfield  {author} {\bibinfo {author} {\bibfnamefont {Z.~A.~H.}\
  \bibnamefont {Goodwin}}, \bibinfo {author} {\bibfnamefont {J.~P.}\
  \bibnamefont {de~Souza}}, \bibinfo {author} {\bibfnamefont {M.~Z.}\
  \bibnamefont {Bazant}}, \ and\ \bibinfo {author} {\bibfnamefont {A.~A.}\
  \bibnamefont {Kornyshev}},\ }\href
  {https://doi.org/10.1007/978-981-10-6739-6_62-1} {\emph {\bibinfo {title}
  {Mean-Field Theory of the Electrical Double Layer in Ionic Liquids}}}\
  (\bibinfo  {publisher} {In: Zhang S. (eds) Encyclopedia of Ionic Liquids.
  Springer, Singapore.},\ \bibinfo {year} {2021})\BibitemShut {NoStop}%
\bibitem [{\citenamefont {Gebbie}\ \emph {et~al.}(2013)\citenamefont {Gebbie},
  \citenamefont {Valtiner}, \citenamefont {Banquy}, \citenamefont {Fox},
  \citenamefont {Henderson},\ and\ \citenamefont {Israelachvili}}]{Gebbie2013}%
  \BibitemOpen
  \bibfield  {author} {\bibinfo {author} {\bibfnamefont {M.~A.}\ \bibnamefont
  {Gebbie}}, \bibinfo {author} {\bibfnamefont {M.}~\bibnamefont {Valtiner}},
  \bibinfo {author} {\bibfnamefont {X.}~\bibnamefont {Banquy}}, \bibinfo
  {author} {\bibfnamefont {E.~T.}\ \bibnamefont {Fox}}, \bibinfo {author}
  {\bibfnamefont {W.~A.}\ \bibnamefont {Henderson}}, \ and\ \bibinfo {author}
  {\bibfnamefont {J.~N.}\ \bibnamefont {Israelachvili}},\ }\href {\doibase
  10.1073/pnas.1307871110} {\bibfield  {journal} {\bibinfo  {journal} {PNAS}\
  }\textbf {\bibinfo {volume} {110}},\ \bibinfo {pages} {9674} (\bibinfo {year}
  {2013})}\BibitemShut {NoStop}%
\bibitem [{\citenamefont {Gebbie}\ \emph {et~al.}(2015)\citenamefont {Gebbie},
  \citenamefont {Dobes}, \citenamefont {Valtiner},\ and\ \citenamefont
  {Israelachvili}}]{Gebbie2015}%
  \BibitemOpen
  \bibfield  {author} {\bibinfo {author} {\bibfnamefont {M.~A.}\ \bibnamefont
  {Gebbie}}, \bibinfo {author} {\bibfnamefont {H.~A.}\ \bibnamefont {Dobes}},
  \bibinfo {author} {\bibfnamefont {M.}~\bibnamefont {Valtiner}}, \ and\
  \bibinfo {author} {\bibfnamefont {J.~N.}\ \bibnamefont {Israelachvili}},\
  }\href@noop {} {\bibfield  {journal} {\bibinfo  {journal} {PNAS}\ }\textbf
  {\bibinfo {volume} {112}},\ \bibinfo {pages} {7432–7437} (\bibinfo {year}
  {2015})}\BibitemShut {NoStop}%
\bibitem [{\citenamefont {Smith}\ \emph {et~al.}(2016)\citenamefont {Smith},
  \citenamefont {Lee},\ and\ \citenamefont {Perkin}}]{Smith2016}%
  \BibitemOpen
  \bibfield  {author} {\bibinfo {author} {\bibfnamefont {A.~M.}\ \bibnamefont
  {Smith}}, \bibinfo {author} {\bibfnamefont {A.~A.}\ \bibnamefont {Lee}}, \
  and\ \bibinfo {author} {\bibfnamefont {S.}~\bibnamefont {Perkin}},\ }\href
  {\doibase 10.1021/acs.jpclett.6b00867} {\bibfield  {journal} {\bibinfo
  {journal} {J. Phys. Chem. Lett.}\ }\textbf {\bibinfo {volume} {7}},\ \bibinfo
  {pages} {2157} (\bibinfo {year} {2016})}\BibitemShut {NoStop}%
\bibitem [{\citenamefont {Gebbie}\ \emph {et~al.}(2017)\citenamefont {Gebbie},
  \citenamefont {Smith}, \citenamefont {Dobbs}, \citenamefont {Lee},
  \citenamefont {Warr}, \citenamefont {Banquy}, \citenamefont {Valtiner},
  \citenamefont {Rutland}, \citenamefont {Israelachvili}, \citenamefont
  {Perkin},\ and\ \citenamefont {Atkin}}]{Gebbie2017rev}%
  \BibitemOpen
  \bibfield  {author} {\bibinfo {author} {\bibfnamefont {M.~A.}\ \bibnamefont
  {Gebbie}}, \bibinfo {author} {\bibfnamefont {A.~M.}\ \bibnamefont {Smith}},
  \bibinfo {author} {\bibfnamefont {H.~A.}\ \bibnamefont {Dobbs}}, \bibinfo
  {author} {\bibfnamefont {A.}~\bibnamefont {Lee}}, \bibinfo {author}
  {\bibfnamefont {G.~G.}\ \bibnamefont {Warr}}, \bibinfo {author}
  {\bibfnamefont {X.}~\bibnamefont {Banquy}}, \bibinfo {author} {\bibfnamefont
  {M.}~\bibnamefont {Valtiner}}, \bibinfo {author} {\bibfnamefont {M.~W.}\
  \bibnamefont {Rutland}}, \bibinfo {author} {\bibfnamefont {J.~N.}\
  \bibnamefont {Israelachvili}}, \bibinfo {author} {\bibfnamefont
  {S.}~\bibnamefont {Perkin}}, \ and\ \bibinfo {author} {\bibfnamefont
  {R.}~\bibnamefont {Atkin}},\ }\href@noop {} {\bibfield  {journal} {\bibinfo
  {journal} {Chem. Commun.}\ }\textbf {\bibinfo {volume} {53}},\ \bibinfo
  {pages} {1214} (\bibinfo {year} {2017})}\BibitemShut {NoStop}%
\bibitem [{\citenamefont {Smith}\ \emph {et~al.}(2017)\citenamefont {Smith},
  \citenamefont {Lee},\ and\ \citenamefont {Perkin}}]{smith2017struct}%
  \BibitemOpen
  \bibfield  {author} {\bibinfo {author} {\bibfnamefont {A.~M.}\ \bibnamefont
  {Smith}}, \bibinfo {author} {\bibfnamefont {A.~A.}\ \bibnamefont {Lee}}, \
  and\ \bibinfo {author} {\bibfnamefont {S.}~\bibnamefont {Perkin}},\
  }\href@noop {} {\bibfield  {journal} {\bibinfo  {journal} {Phys. Rev. Lett.}\
  }\textbf {\bibinfo {volume} {118}},\ \bibinfo {pages} {096002} (\bibinfo
  {year} {2017})}\BibitemShut {NoStop}%
\bibitem [{\citenamefont {Han}\ \emph {et~al.}(2020)\citenamefont {Han},
  \citenamefont {Kim}, \citenamefont {Leal}, \citenamefont {Negrito},
  \citenamefont {Batteas},\ and\ \citenamefont {Espinosa-Marzal}}]{Han2020IL}%
  \BibitemOpen
  \bibfield  {author} {\bibinfo {author} {\bibfnamefont {M.}~\bibnamefont
  {Han}}, \bibinfo {author} {\bibfnamefont {H.}~\bibnamefont {Kim}}, \bibinfo
  {author} {\bibfnamefont {C.}~\bibnamefont {Leal}}, \bibinfo {author}
  {\bibfnamefont {M.}~\bibnamefont {Negrito}}, \bibinfo {author} {\bibfnamefont
  {J.~D.}\ \bibnamefont {Batteas}}, \ and\ \bibinfo {author} {\bibfnamefont
  {R.~M.}\ \bibnamefont {Espinosa-Marzal}},\ }\href@noop {} {\bibfield
  {journal} {\bibinfo  {journal} {Adv. Mater.}\ }\textbf {\bibinfo {volume}
  {7}},\ \bibinfo {pages} {2001313} (\bibinfo {year} {2020})}\BibitemShut
  {NoStop}%
\bibitem [{\citenamefont {Jurado}\ \emph {et~al.}(2016)\citenamefont {Jurado},
  \citenamefont {Kim}, \citenamefont {Rossi}, \citenamefont {Arcifa},
  \citenamefont {Schuh}, \citenamefont {Spencer}, \citenamefont {Leal},
  \citenamefont {Ewoldte},\ and\ \citenamefont {Espinosa-Marzal}}]{Jurado2016}%
  \BibitemOpen
  \bibfield  {author} {\bibinfo {author} {\bibfnamefont {L.~A.}\ \bibnamefont
  {Jurado}}, \bibinfo {author} {\bibfnamefont {H.}~\bibnamefont {Kim}},
  \bibinfo {author} {\bibfnamefont {A.}~\bibnamefont {Rossi}}, \bibinfo
  {author} {\bibfnamefont {A.}~\bibnamefont {Arcifa}}, \bibinfo {author}
  {\bibfnamefont {J.~K.}\ \bibnamefont {Schuh}}, \bibinfo {author}
  {\bibfnamefont {N.~D.}\ \bibnamefont {Spencer}}, \bibinfo {author}
  {\bibfnamefont {C.}~\bibnamefont {Leal}}, \bibinfo {author} {\bibfnamefont
  {R.~H.}\ \bibnamefont {Ewoldte}}, \ and\ \bibinfo {author} {\bibfnamefont
  {R.~M.}\ \bibnamefont {Espinosa-Marzal}},\ }\href@noop {} {\bibfield
  {journal} {\bibinfo  {journal} {Phys. Chem. Chem. Phys.}\ }\textbf {\bibinfo
  {volume} {18}},\ \bibinfo {pages} {22719} (\bibinfo {year}
  {2016})}\BibitemShut {NoStop}%
\bibitem [{\citenamefont {Jurado}\ \emph {et~al.}(2015)\citenamefont {Jurado},
  \citenamefont {Kim}, \citenamefont {Arcifa}, \citenamefont {Rossi},
  \citenamefont {Leal}, \citenamefont {Spencerc},\ and\ \citenamefont
  {Espinosa-Marzal}}]{Jurado2015}%
  \BibitemOpen
  \bibfield  {author} {\bibinfo {author} {\bibfnamefont {L.~A.}\ \bibnamefont
  {Jurado}}, \bibinfo {author} {\bibfnamefont {H.}~\bibnamefont {Kim}},
  \bibinfo {author} {\bibfnamefont {A.}~\bibnamefont {Arcifa}}, \bibinfo
  {author} {\bibfnamefont {A.}~\bibnamefont {Rossi}}, \bibinfo {author}
  {\bibfnamefont {C.}~\bibnamefont {Leal}}, \bibinfo {author} {\bibfnamefont
  {N.~D.}\ \bibnamefont {Spencerc}}, \ and\ \bibinfo {author} {\bibfnamefont
  {R.~M.}\ \bibnamefont {Espinosa-Marzal}},\ }\href@noop {} {\bibfield
  {journal} {\bibinfo  {journal} {Phys. Chem. Chem. Phys.}\ }\textbf {\bibinfo
  {volume} {17}},\ \bibinfo {pages} {13613} (\bibinfo {year}
  {2015})}\BibitemShut {NoStop}%
\bibitem [{\citenamefont {Jurado}\ and\ \citenamefont
  {Espinosa-Marzal}(2017)}]{Jurado2017EDL}%
  \BibitemOpen
  \bibfield  {author} {\bibinfo {author} {\bibfnamefont {L.~A.}\ \bibnamefont
  {Jurado}}\ and\ \bibinfo {author} {\bibfnamefont {R.~M.}\ \bibnamefont
  {Espinosa-Marzal}},\ }\href@noop {} {\bibfield  {journal} {\bibinfo
  {journal} {Sci. Rep.}\ }\textbf {\bibinfo {volume} {17}},\ \bibinfo {pages}
  {4225} (\bibinfo {year} {2017})}\BibitemShut {NoStop}%
\bibitem [{\citenamefont {Mao}\ \emph {et~al.}(2019)\citenamefont {Mao},
  \citenamefont {Brown}, \citenamefont {\v{C}ervinka}, \citenamefont {Hazell},
  \citenamefont {Li}, \citenamefont {Ren}, \citenamefont {Chen}, \citenamefont
  {Atkin}, \citenamefont {Eastoe}, \citenamefont {Grillo}, \citenamefont
  {Padua}, \citenamefont {Gomes},\ and\ \citenamefont {Hatton}}]{Mao2019nano}%
  \BibitemOpen
  \bibfield  {author} {\bibinfo {author} {\bibfnamefont {X.}~\bibnamefont
  {Mao}}, \bibinfo {author} {\bibfnamefont {P.}~\bibnamefont {Brown}}, \bibinfo
  {author} {\bibfnamefont {C.}~\bibnamefont {\v{C}ervinka}}, \bibinfo {author}
  {\bibfnamefont {G.}~\bibnamefont {Hazell}}, \bibinfo {author} {\bibfnamefont
  {H.}~\bibnamefont {Li}}, \bibinfo {author} {\bibfnamefont {Y.}~\bibnamefont
  {Ren}}, \bibinfo {author} {\bibfnamefont {D.}~\bibnamefont {Chen}}, \bibinfo
  {author} {\bibfnamefont {R.}~\bibnamefont {Atkin}}, \bibinfo {author}
  {\bibfnamefont {J.}~\bibnamefont {Eastoe}}, \bibinfo {author} {\bibfnamefont
  {I.}~\bibnamefont {Grillo}}, \bibinfo {author} {\bibfnamefont {A.~A.~H.}\
  \bibnamefont {Padua}}, \bibinfo {author} {\bibfnamefont {M.~F.~C.}\
  \bibnamefont {Gomes}}, \ and\ \bibinfo {author} {\bibfnamefont {T.~A.}\
  \bibnamefont {Hatton}},\ }\href@noop {} {\bibfield  {journal} {\bibinfo
  {journal} {Nat. Mater.}\ }\textbf {\bibinfo {volume} {18}},\ \bibinfo {pages}
  {1350} (\bibinfo {year} {2019})}\BibitemShut {NoStop}%
\bibitem [{\citenamefont {Hjalmarsson}\ \emph {et~al.}(2017)\citenamefont
  {Hjalmarsson}, \citenamefont {Atkin},\ and\ \citenamefont
  {Rutland}}]{Hjalmarsson2017}%
  \BibitemOpen
  \bibfield  {author} {\bibinfo {author} {\bibfnamefont {N.}~\bibnamefont
  {Hjalmarsson}}, \bibinfo {author} {\bibfnamefont {R.}~\bibnamefont {Atkin}},
  \ and\ \bibinfo {author} {\bibfnamefont {M.~W.}\ \bibnamefont {Rutland}},\
  }\href@noop {} {\bibfield  {journal} {\bibinfo  {journal} {Chem. Commun.}\
  }\textbf {\bibinfo {volume} {53}},\ \bibinfo {pages} {647} (\bibinfo {year}
  {2017})}\BibitemShut {NoStop}%
\bibitem [{\citenamefont {Comtet}\ \emph {et~al.}(2017)\citenamefont {Comtet},
  \citenamefont {Nigu\`es}, \citenamefont {Kaiser}, \citenamefont {Coasne},
  \citenamefont {Bocquet},\ and\ \citenamefont {Siria}}]{comtet2017nano}%
  \BibitemOpen
  \bibfield  {author} {\bibinfo {author} {\bibfnamefont {J.}~\bibnamefont
  {Comtet}}, \bibinfo {author} {\bibfnamefont {A.}~\bibnamefont {Nigu\`es}},
  \bibinfo {author} {\bibfnamefont {V.}~\bibnamefont {Kaiser}}, \bibinfo
  {author} {\bibfnamefont {B.}~\bibnamefont {Coasne}}, \bibinfo {author}
  {\bibfnamefont {L.}~\bibnamefont {Bocquet}}, \ and\ \bibinfo {author}
  {\bibfnamefont {A.}~\bibnamefont {Siria}},\ }\href@noop {} {\bibfield
  {journal} {\bibinfo  {journal} {Nat. Mater.}\ }\textbf {\bibinfo {volume}
  {16}},\ \bibinfo {pages} {634} (\bibinfo {year} {2017})}\BibitemShut
  {NoStop}%
\bibitem [{\citenamefont {Fedorov}\ and\ \citenamefont
  {Kornyshev}(2008{\natexlab{a}})}]{Fedorov2008a}%
  \BibitemOpen
  \bibfield  {author} {\bibinfo {author} {\bibfnamefont {M.~V.}\ \bibnamefont
  {Fedorov}}\ and\ \bibinfo {author} {\bibfnamefont {A.~A.}\ \bibnamefont
  {Kornyshev}},\ }\href {\doibase 10.1021/jp803440q} {\bibfield  {journal}
  {\bibinfo  {journal} {J. Phys. Chem. B}\ }\textbf {\bibinfo {volume} {112}},\
  \bibinfo {pages} {11868} (\bibinfo {year} {2008}{\natexlab{a}})}\BibitemShut
  {NoStop}%
\bibitem [{\citenamefont {Fedorov}\ and\ \citenamefont
  {Kornyshev}(2008{\natexlab{b}})}]{Fedorov2008}%
  \BibitemOpen
  \bibfield  {author} {\bibinfo {author} {\bibfnamefont {M.~V.}\ \bibnamefont
  {Fedorov}}\ and\ \bibinfo {author} {\bibfnamefont {A.~A.}\ \bibnamefont
  {Kornyshev}},\ }\href@noop {} {\bibfield  {journal} {\bibinfo  {journal}
  {Electrochim. Acta}\ }\textbf {\bibinfo {volume} {53}},\ \bibinfo {pages}
  {6835} (\bibinfo {year} {2008}{\natexlab{b}})}\BibitemShut {NoStop}%
\bibitem [{\citenamefont {Georgi}\ \emph {et~al.}(2010)\citenamefont {Georgi},
  \citenamefont {Kornyshev},\ and\ \citenamefont {Fedorov}}]{Georgi2010}%
  \BibitemOpen
  \bibfield  {author} {\bibinfo {author} {\bibfnamefont {N.}~\bibnamefont
  {Georgi}}, \bibinfo {author} {\bibfnamefont {A.}~\bibnamefont {Kornyshev}}, \
  and\ \bibinfo {author} {\bibfnamefont {M.}~\bibnamefont {Fedorov}},\ }\href
  {\doibase 10.1016/j.jelechem.2010.07.004} {\bibfield  {journal} {\bibinfo
  {journal} {Journal of Electroanalytical Chemistry}\ }\textbf {\bibinfo
  {volume} {649}},\ \bibinfo {pages} {261} (\bibinfo {year}
  {2010})}\BibitemShut {NoStop}%
\bibitem [{\citenamefont {Bazant}\ \emph {et~al.}(2011)\citenamefont {Bazant},
  \citenamefont {Storey},\ and\ \citenamefont {Kornyshev}}]{Bazant2011}%
  \BibitemOpen
  \bibfield  {author} {\bibinfo {author} {\bibfnamefont {M.~Z.}\ \bibnamefont
  {Bazant}}, \bibinfo {author} {\bibfnamefont {B.~D.}\ \bibnamefont {Storey}},
  \ and\ \bibinfo {author} {\bibfnamefont {A.~A.}\ \bibnamefont {Kornyshev}},\
  }\href {\doibase 10.1103/PhysRevLett.106.046102} {\bibfield  {journal}
  {\bibinfo  {journal} {Phys. Rev. Lett.}\ }\textbf {\bibinfo {volume} {106}},\
  \bibinfo {pages} {046102} (\bibinfo {year} {2011})}\BibitemShut {NoStop}%
\bibitem [{\citenamefont {Coles}\ \emph {et~al.}(2020)\citenamefont {Coles},
  \citenamefont {Park}, \citenamefont {Nikam}, \citenamefont {Kanduc},
  \citenamefont {Dzubiella},\ and\ \citenamefont
  {Rotenberg}}]{Coles2020length}%
  \BibitemOpen
  \bibfield  {author} {\bibinfo {author} {\bibfnamefont {S.}~\bibnamefont
  {Coles}}, \bibinfo {author} {\bibfnamefont {C.}~\bibnamefont {Park}},
  \bibinfo {author} {\bibfnamefont {R.}~\bibnamefont {Nikam}}, \bibinfo
  {author} {\bibfnamefont {M.}~\bibnamefont {Kanduc}}, \bibinfo {author}
  {\bibfnamefont {J.}~\bibnamefont {Dzubiella}}, \ and\ \bibinfo {author}
  {\bibfnamefont {B.}~\bibnamefont {Rotenberg}},\ }\href@noop {} {\bibfield
  {journal} {\bibinfo  {journal} {J. Phys. Chem. B}\ }\textbf {\bibinfo
  {volume} {124}},\ \bibinfo {pages} {1778} (\bibinfo {year}
  {2020})}\BibitemShut {NoStop}%
\bibitem [{\citenamefont {de~Souza}\ \emph {et~al.}(2020)\citenamefont
  {de~Souza}, \citenamefont {Goodwin}, \citenamefont {McEldrew}, \citenamefont
  {Kornyshev},\ and\ \citenamefont {Bazant}}]{pedroRTILs}%
  \BibitemOpen
  \bibfield  {author} {\bibinfo {author} {\bibfnamefont {J.~P.}\ \bibnamefont
  {de~Souza}}, \bibinfo {author} {\bibfnamefont {Z.~A.~H.}\ \bibnamefont
  {Goodwin}}, \bibinfo {author} {\bibfnamefont {M.}~\bibnamefont {McEldrew}},
  \bibinfo {author} {\bibfnamefont {A.~A.}\ \bibnamefont {Kornyshev}}, \ and\
  \bibinfo {author} {\bibfnamefont {M.~Z.}\ \bibnamefont {Bazant}},\ }\href
  {\doibase 10.1103/PhysRevLett.125.116001} {\bibfield  {journal} {\bibinfo
  {journal} {Phys. Rev. Lett.}\ }\textbf {\bibinfo {volume} {125}},\ \bibinfo
  {pages} {116001} (\bibinfo {year} {2020})}\BibitemShut {NoStop}%
\bibitem [{\citenamefont {de~Souza}\ \emph {et~al.}(2021)\citenamefont
  {de~Souza}, \citenamefont {Pivnic}, \citenamefont {Bazant}, \citenamefont
  {Urbakh},\ and\ \citenamefont {Kornyshev}}]{pedro2022force}%
  \BibitemOpen
  \bibfield  {author} {\bibinfo {author} {\bibfnamefont {J.~P.}\ \bibnamefont
  {de~Souza}}, \bibinfo {author} {\bibfnamefont {K.}~\bibnamefont {Pivnic}},
  \bibinfo {author} {\bibfnamefont {M.~Z.}\ \bibnamefont {Bazant}}, \bibinfo
  {author} {\bibfnamefont {M.}~\bibnamefont {Urbakh}}, \ and\ \bibinfo {author}
  {\bibfnamefont {A.~A.}\ \bibnamefont {Kornyshev}},\ }\href@noop {} {\bibfield
   {journal} {\bibinfo  {journal} {J. Phys. Chem. B}\ }\textbf {\bibinfo
  {volume} {126}},\ \bibinfo {pages} {1242} (\bibinfo {year}
  {2021})}\BibitemShut {NoStop}%
\bibitem [{\citenamefont {Gavish}\ \emph {et~al.}(2018)\citenamefont {Gavish},
  \citenamefont {Elad},\ and\ \citenamefont {Yochelis}}]{gavish2018solvent}%
  \BibitemOpen
  \bibfield  {author} {\bibinfo {author} {\bibfnamefont {N.}~\bibnamefont
  {Gavish}}, \bibinfo {author} {\bibfnamefont {D.}~\bibnamefont {Elad}}, \ and\
  \bibinfo {author} {\bibfnamefont {A.}~\bibnamefont {Yochelis}},\ }\href@noop
  {} {\bibfield  {journal} {\bibinfo  {journal} {J. Phys. Chem. Lett.}\
  }\textbf {\bibinfo {volume} {9}},\ \bibinfo {pages} {36} (\bibinfo {year}
  {2018})}\BibitemShut {NoStop}%
\bibitem [{\citenamefont {Krucker-Velasquez}\ and\ \citenamefont
  {Swan}(2021)}]{emily2021}%
  \BibitemOpen
  \bibfield  {author} {\bibinfo {author} {\bibfnamefont {E.}~\bibnamefont
  {Krucker-Velasquez}}\ and\ \bibinfo {author} {\bibfnamefont {J.~W.}\
  \bibnamefont {Swan}},\ }\href@noop {} {\bibfield  {journal} {\bibinfo
  {journal} {J. Chem. Phys.}\ }\textbf {\bibinfo {volume} {155}},\ \bibinfo
  {pages} {134903} (\bibinfo {year} {2021})}\BibitemShut {NoStop}%
\bibitem [{\citenamefont {Levy}\ \emph {et~al.}(2019)\citenamefont {Levy},
  \citenamefont {McEldrew},\ and\ \citenamefont {Bazant}}]{levy2019spin}%
  \BibitemOpen
  \bibfield  {author} {\bibinfo {author} {\bibfnamefont {A.}~\bibnamefont
  {Levy}}, \bibinfo {author} {\bibfnamefont {M.}~\bibnamefont {McEldrew}}, \
  and\ \bibinfo {author} {\bibfnamefont {M.~Z.}\ \bibnamefont {Bazant}},\
  }\href@noop {} {\bibfield  {journal} {\bibinfo  {journal} {Phys. Rev.
  Mater.}\ }\textbf {\bibinfo {volume} {3}},\ \bibinfo {pages} {055606}
  (\bibinfo {year} {2019})}\BibitemShut {NoStop}%
\bibitem [{\citenamefont {Zhao}\ \emph {et~al.}(2009)\citenamefont {Zhao},
  \citenamefont {Leroy}, \citenamefont {Heggen}, \citenamefont {Zahn},
  \citenamefont {Kirchner}, \citenamefont {Balasubramanian},\ and\
  \citenamefont {M\"{u}ller-Plathe}}]{Zhao2009}%
  \BibitemOpen
  \bibfield  {author} {\bibinfo {author} {\bibfnamefont {W.}~\bibnamefont
  {Zhao}}, \bibinfo {author} {\bibfnamefont {F.}~\bibnamefont {Leroy}},
  \bibinfo {author} {\bibfnamefont {B.}~\bibnamefont {Heggen}}, \bibinfo
  {author} {\bibfnamefont {S.}~\bibnamefont {Zahn}}, \bibinfo {author}
  {\bibfnamefont {B.}~\bibnamefont {Kirchner}}, \bibinfo {author}
  {\bibfnamefont {S.}~\bibnamefont {Balasubramanian}}, \ and\ \bibinfo {author}
  {\bibfnamefont {F.}~\bibnamefont {M\"{u}ller-Plathe}},\ }\href@noop {}
  {\bibfield  {journal} {\bibinfo  {journal} {J. AM. CHEM. SOC.}\ }\textbf
  {\bibinfo {volume} {131}},\ \bibinfo {pages} {15825} (\bibinfo {year}
  {2009})}\BibitemShut {NoStop}%
\bibitem [{\citenamefont {Lee}\ \emph {et~al.}(2015)\citenamefont {Lee},
  \citenamefont {Vella}, \citenamefont {Perkin},\ and\ \citenamefont
  {Goriely}}]{Lee2015}%
  \BibitemOpen
  \bibfield  {author} {\bibinfo {author} {\bibfnamefont {A.~A.}\ \bibnamefont
  {Lee}}, \bibinfo {author} {\bibfnamefont {D.}~\bibnamefont {Vella}}, \bibinfo
  {author} {\bibfnamefont {S.}~\bibnamefont {Perkin}}, \ and\ \bibinfo {author}
  {\bibfnamefont {A.}~\bibnamefont {Goriely}},\ }\href {\doibase
  10.1021/jz502250z} {\bibfield  {journal} {\bibinfo  {journal} {J. Phys. Chem.
  Lett.}\ }\textbf {\bibinfo {volume} {6}},\ \bibinfo {pages} {159} (\bibinfo
  {year} {2015})}\BibitemShut {NoStop}%
\bibitem [{\citenamefont {Zhang}\ and\ \citenamefont
  {Maginn}(2015)}]{Zhang2015}%
  \BibitemOpen
  \bibfield  {author} {\bibinfo {author} {\bibfnamefont {Y.}~\bibnamefont
  {Zhang}}\ and\ \bibinfo {author} {\bibfnamefont {E.~J.}\ \bibnamefont
  {Maginn}},\ }\href@noop {} {\bibfield  {journal} {\bibinfo  {journal} {J.
  Phys. Chem. Lett.}\ }\textbf {\bibinfo {volume} {6}},\ \bibinfo {pages} {700}
  (\bibinfo {year} {2015})}\BibitemShut {NoStop}%
\bibitem [{\citenamefont {Holl{\'o}czki}\ \emph {et~al.}(2014)\citenamefont
  {Holl{\'o}czki}, \citenamefont {Malberg}, \citenamefont {Welton},\ and\
  \citenamefont {Kirchner}}]{Kirchner2014}%
  \BibitemOpen
  \bibfield  {author} {\bibinfo {author} {\bibfnamefont {O.}~\bibnamefont
  {Holl{\'o}czki}}, \bibinfo {author} {\bibfnamefont {F.}~\bibnamefont
  {Malberg}}, \bibinfo {author} {\bibfnamefont {T.}~\bibnamefont {Welton}}, \
  and\ \bibinfo {author} {\bibfnamefont {B.}~\bibnamefont {Kirchner}},\
  }\href@noop {} {\bibfield  {journal} {\bibinfo  {journal} {Phys. Chem. Chem.
  Phys.}\ }\textbf {\bibinfo {volume} {16}},\ \bibinfo {pages} {16880}
  (\bibinfo {year} {2014})}\BibitemShut {NoStop}%
\bibitem [{\citenamefont {Araque}\ \emph {et~al.}(2015)\citenamefont {Araque},
  \citenamefont {Yadav}, \citenamefont {Shadeck}, \citenamefont {Maroncelli},\
  and\ \citenamefont {Margulis}}]{Araque2015}%
  \BibitemOpen
  \bibfield  {author} {\bibinfo {author} {\bibfnamefont {J.~C.}\ \bibnamefont
  {Araque}}, \bibinfo {author} {\bibfnamefont {S.~K.}\ \bibnamefont {Yadav}},
  \bibinfo {author} {\bibfnamefont {M.}~\bibnamefont {Shadeck}}, \bibinfo
  {author} {\bibfnamefont {M.}~\bibnamefont {Maroncelli}}, \ and\ \bibinfo
  {author} {\bibfnamefont {C.~J.}\ \bibnamefont {Margulis}},\ }\href@noop {}
  {\bibfield  {journal} {\bibinfo  {journal} {J. Phys. Chem. B}\ }\textbf
  {\bibinfo {volume} {119}},\ \bibinfo {pages} {7015} (\bibinfo {year}
  {2015})}\BibitemShut {NoStop}%
\bibitem [{\citenamefont {Kirchner}\ \emph {et~al.}(2015)\citenamefont
  {Kirchner}, \citenamefont {Malberg}, \citenamefont {Firaha},\ and\
  \citenamefont {Holl\'oczki}}]{Krichner2015ionpair}%
  \BibitemOpen
  \bibfield  {author} {\bibinfo {author} {\bibfnamefont {B.}~\bibnamefont
  {Kirchner}}, \bibinfo {author} {\bibfnamefont {F.}~\bibnamefont {Malberg}},
  \bibinfo {author} {\bibfnamefont {D.~S.}\ \bibnamefont {Firaha}}, \ and\
  \bibinfo {author} {\bibfnamefont {O.}~\bibnamefont {Holl\'oczki}},\
  }\href@noop {} {\bibfield  {journal} {\bibinfo  {journal} {J. Phys.: Condens.
  Matter}\ }\textbf {\bibinfo {volume} {27}},\ \bibinfo {pages} {463002}
  (\bibinfo {year} {2015})}\BibitemShut {NoStop}%
\bibitem [{\citenamefont {Ma}\ \emph {et~al.}(2015)\citenamefont {Ma},
  \citenamefont {Forsman},\ and\ \citenamefont {Woodward}}]{Ma2015}%
  \BibitemOpen
  \bibfield  {author} {\bibinfo {author} {\bibfnamefont {K.}~\bibnamefont
  {Ma}}, \bibinfo {author} {\bibfnamefont {J.}~\bibnamefont {Forsman}}, \ and\
  \bibinfo {author} {\bibfnamefont {C.~E.}\ \bibnamefont {Woodward}},\
  }\href@noop {} {\bibfield  {journal} {\bibinfo  {journal} {J. Chem. Phys.}\
  }\textbf {\bibinfo {volume} {142}},\ \bibinfo {pages} {174704} (\bibinfo
  {year} {2015})}\BibitemShut {NoStop}%
\bibitem [{\citenamefont {Adar}\ \emph {et~al.}(2017)\citenamefont {Adar},
  \citenamefont {Markovich},\ and\ \citenamefont {Andelman}}]{adar2017bjerrum}%
  \BibitemOpen
  \bibfield  {author} {\bibinfo {author} {\bibfnamefont {R.~M.}\ \bibnamefont
  {Adar}}, \bibinfo {author} {\bibfnamefont {T.}~\bibnamefont {Markovich}}, \
  and\ \bibinfo {author} {\bibfnamefont {D.}~\bibnamefont {Andelman}},\
  }\href@noop {} {\bibfield  {journal} {\bibinfo  {journal} {J. Chem. Phys.}\
  }\textbf {\bibinfo {volume} {146}},\ \bibinfo {pages} {194904} (\bibinfo
  {year} {2017})}\BibitemShut {NoStop}%
\bibitem [{\citenamefont {Zhang}\ \emph {et~al.}(2020)\citenamefont {Zhang},
  \citenamefont {Ye}, \citenamefont {Chen}, \citenamefont {Goodwin},
  \citenamefont {Feng}, \citenamefont {Huang},\ and\ \citenamefont
  {Kornyshev}}]{yufan2020}%
  \BibitemOpen
  \bibfield  {author} {\bibinfo {author} {\bibfnamefont {Y.}~\bibnamefont
  {Zhang}}, \bibinfo {author} {\bibfnamefont {T.}~\bibnamefont {Ye}}, \bibinfo
  {author} {\bibfnamefont {M.}~\bibnamefont {Chen}}, \bibinfo {author}
  {\bibfnamefont {Z.~A.~H.}\ \bibnamefont {Goodwin}}, \bibinfo {author}
  {\bibfnamefont {G.}~\bibnamefont {Feng}}, \bibinfo {author} {\bibfnamefont
  {J.}~\bibnamefont {Huang}}, \ and\ \bibinfo {author} {\bibfnamefont {A.~A.}\
  \bibnamefont {Kornyshev}},\ }\href@noop {} {\bibfield  {journal} {\bibinfo
  {journal} {Energy Environ. Mater.}\ }\textbf {\bibinfo {volume} {3}},\
  \bibinfo {pages} {414} (\bibinfo {year} {2020})}\BibitemShut {NoStop}%
\bibitem [{\citenamefont {Feng}\ \emph {et~al.}(2019)\citenamefont {Feng},
  \citenamefont {Chen}, \citenamefont {Bi}, \citenamefont {Goodwin},
  \citenamefont {Postnikov}, \citenamefont {Brilliantov}, \citenamefont
  {Urbakh},\ and\ \citenamefont {Kornyshev}}]{feng2019free}%
  \BibitemOpen
  \bibfield  {author} {\bibinfo {author} {\bibfnamefont {G.}~\bibnamefont
  {Feng}}, \bibinfo {author} {\bibfnamefont {M.}~\bibnamefont {Chen}}, \bibinfo
  {author} {\bibfnamefont {S.}~\bibnamefont {Bi}}, \bibinfo {author}
  {\bibfnamefont {Z.~A.}\ \bibnamefont {Goodwin}}, \bibinfo {author}
  {\bibfnamefont {E.~B.}\ \bibnamefont {Postnikov}}, \bibinfo {author}
  {\bibfnamefont {N.}~\bibnamefont {Brilliantov}}, \bibinfo {author}
  {\bibfnamefont {M.}~\bibnamefont {Urbakh}}, \ and\ \bibinfo {author}
  {\bibfnamefont {A.~A.}\ \bibnamefont {Kornyshev}},\ }\href@noop {} {\bibfield
   {journal} {\bibinfo  {journal} {Phys. Rev. X}\ }\textbf {\bibinfo {volume}
  {9}},\ \bibinfo {pages} {021024} (\bibinfo {year} {2019})}\BibitemShut
  {NoStop}%
\bibitem [{\citenamefont {Avni}\ \emph {et~al.}(2020)\citenamefont {Avni},
  \citenamefont {Adar},\ and\ \citenamefont {Andelman}}]{avni2020charge}%
  \BibitemOpen
  \bibfield  {author} {\bibinfo {author} {\bibfnamefont {Y.}~\bibnamefont
  {Avni}}, \bibinfo {author} {\bibfnamefont {R.~M.}\ \bibnamefont {Adar}}, \
  and\ \bibinfo {author} {\bibfnamefont {D.}~\bibnamefont {Andelman}},\
  }\href@noop {} {\bibfield  {journal} {\bibinfo  {journal} {Phys. Rev. E}\
  }\textbf {\bibinfo {volume} {101}},\ \bibinfo {pages} {010601} (\bibinfo
  {year} {2020})}\BibitemShut {NoStop}%
\bibitem [{\citenamefont {Chen}\ \emph {et~al.}(2018)\citenamefont {Chen},
  \citenamefont {Goodwin}, \citenamefont {Feng},\ and\ \citenamefont
  {Kornyshev}}]{Chen2017}%
  \BibitemOpen
  \bibfield  {author} {\bibinfo {author} {\bibfnamefont {M.}~\bibnamefont
  {Chen}}, \bibinfo {author} {\bibfnamefont {Z.~A.~H.}\ \bibnamefont
  {Goodwin}}, \bibinfo {author} {\bibfnamefont {G.}~\bibnamefont {Feng}}, \
  and\ \bibinfo {author} {\bibfnamefont {A.~A.}\ \bibnamefont {Kornyshev}},\
  }\href {\doibase 10.1016/j.jelechem.2017.11.005} {\bibfield  {journal}
  {\bibinfo  {journal} {J. Electroanal. Chem.}\ }\textbf {\bibinfo {volume}
  {819}},\ \bibinfo {pages} {347} (\bibinfo {year} {2018})}\BibitemShut
  {NoStop}%
\bibitem [{\citenamefont {Goodwin}\ \emph {et~al.}(2017)\citenamefont
  {Goodwin}, \citenamefont {Feng},\ and\ \citenamefont
  {Kornyshev}}]{goodwin2017mean}%
  \BibitemOpen
  \bibfield  {author} {\bibinfo {author} {\bibfnamefont {Z.~A.}\ \bibnamefont
  {Goodwin}}, \bibinfo {author} {\bibfnamefont {G.}~\bibnamefont {Feng}}, \
  and\ \bibinfo {author} {\bibfnamefont {A.~A.}\ \bibnamefont {Kornyshev}},\
  }\href@noop {} {\bibfield  {journal} {\bibinfo  {journal} {Electrochim.
  Acta}\ }\textbf {\bibinfo {volume} {225}},\ \bibinfo {pages} {190} (\bibinfo
  {year} {2017})}\BibitemShut {NoStop}%
\bibitem [{\citenamefont {Goodwin}\ and\ \citenamefont
  {Kornyshev}(2017)}]{goodwin2017underscreening}%
  \BibitemOpen
  \bibfield  {author} {\bibinfo {author} {\bibfnamefont {Z.~A.}\ \bibnamefont
  {Goodwin}}\ and\ \bibinfo {author} {\bibfnamefont {A.~A.}\ \bibnamefont
  {Kornyshev}},\ }\href@noop {} {\bibfield  {journal} {\bibinfo  {journal}
  {Electrochem. commun.}\ }\textbf {\bibinfo {volume} {82}},\ \bibinfo {pages}
  {129} (\bibinfo {year} {2017})}\BibitemShut {NoStop}%
\bibitem [{\citenamefont {Dupont}(2011)}]{Dupont2004}%
  \BibitemOpen
  \bibfield  {author} {\bibinfo {author} {\bibfnamefont {J.}~\bibnamefont
  {Dupont}},\ }\href@noop {} {\bibfield  {journal} {\bibinfo  {journal} {J.
  Braz. Chem. Soc.}\ }\textbf {\bibinfo {volume} {3}},\ \bibinfo {pages} {341}
  (\bibinfo {year} {2011})}\BibitemShut {NoStop}%
\bibitem [{\citenamefont {Singh}\ and\ \citenamefont
  {Kumar}(2008)}]{Singh2008}%
  \BibitemOpen
  \bibfield  {author} {\bibinfo {author} {\bibfnamefont {T.}~\bibnamefont
  {Singh}}\ and\ \bibinfo {author} {\bibfnamefont {A.}~\bibnamefont {Kumar}},\
  }\href@noop {} {\bibfield  {journal} {\bibinfo  {journal} {Colloids and
  Surfaces A: Physicochem. Eng. Aspects}\ }\textbf {\bibinfo {volume} {318}},\
  \bibinfo {pages} {263} (\bibinfo {year} {2008})}\BibitemShut {NoStop}%
\bibitem [{\citenamefont {Dupont}(2004)}]{Dupont2011}%
  \BibitemOpen
  \bibfield  {author} {\bibinfo {author} {\bibfnamefont {J.}~\bibnamefont
  {Dupont}},\ }\href@noop {} {\bibfield  {journal} {\bibinfo  {journal} {Acc.
  Chem. Res.}\ }\textbf {\bibinfo {volume} {44}},\ \bibinfo {pages} {1223}
  (\bibinfo {year} {2004})}\BibitemShut {NoStop}%
\bibitem [{\citenamefont {Wang}\ and\ \citenamefont {Voth}(2005)}]{Wang2005}%
  \BibitemOpen
  \bibfield  {author} {\bibinfo {author} {\bibfnamefont {Y.}~\bibnamefont
  {Wang}}\ and\ \bibinfo {author} {\bibfnamefont {G.~A.}\ \bibnamefont
  {Voth}},\ }\href@noop {} {\bibfield  {journal} {\bibinfo  {journal} {J. Am.
  Chem. Soc.}\ }\textbf {\bibinfo {volume} {35}},\ \bibinfo {pages} {12192}
  (\bibinfo {year} {2005})}\BibitemShut {NoStop}%
\bibitem [{\citenamefont {Hu}\ and\ \citenamefont
  {Margulis}(2006)}]{Hu2006hetero}%
  \BibitemOpen
  \bibfield  {author} {\bibinfo {author} {\bibfnamefont {Z.}~\bibnamefont
  {Hu}}\ and\ \bibinfo {author} {\bibfnamefont {C.~J.}\ \bibnamefont
  {Margulis}},\ }\href@noop {} {\bibfield  {journal} {\bibinfo  {journal}
  {PNAS}\ }\textbf {\bibinfo {volume} {103}},\ \bibinfo {pages} {831} (\bibinfo
  {year} {2006})}\BibitemShut {NoStop}%
\bibitem [{\citenamefont {Bernardes}\ \emph {et~al.}(2011)\citenamefont
  {Bernardes}, \citenamefont {da~Piedade},\ and\ \citenamefont
  {Lopes}}]{Bernardes2011}%
  \BibitemOpen
  \bibfield  {author} {\bibinfo {author} {\bibfnamefont {C.~E.~S.}\
  \bibnamefont {Bernardes}}, \bibinfo {author} {\bibfnamefont {M.~E.~M.}\
  \bibnamefont {da~Piedade}}, \ and\ \bibinfo {author} {\bibfnamefont
  {J.~N.~C.}\ \bibnamefont {Lopes}},\ }\href@noop {} {\bibfield  {journal}
  {\bibinfo  {journal} {J. Phys. Chem. B}\ }\textbf {\bibinfo {volume} {115}},\
  \bibinfo {pages} {2067} (\bibinfo {year} {2011})}\BibitemShut {NoStop}%
\bibitem [{\citenamefont {Lopes}\ and\ \citenamefont
  {P\'adua}(2006)}]{Lopes2006}%
  \BibitemOpen
  \bibfield  {author} {\bibinfo {author} {\bibfnamefont {J.~N. A.~C.}\
  \bibnamefont {Lopes}}\ and\ \bibinfo {author} {\bibfnamefont {A.~A.~H.}\
  \bibnamefont {P\'adua}},\ }\href@noop {} {\bibfield  {journal} {\bibinfo
  {journal} {J. Phys. Chem. B}\ }\textbf {\bibinfo {volume} {110}},\ \bibinfo
  {pages} {3330} (\bibinfo {year} {2006})}\BibitemShut {NoStop}%
\bibitem [{\citenamefont {Borodin}\ \emph {et~al.}(2017)\citenamefont
  {Borodin}, \citenamefont {Suo}, \citenamefont {Gobet}, \citenamefont {Ren},
  \citenamefont {Wang}, \citenamefont {Faraone}, \citenamefont {Peng},
  \citenamefont {Olguin}, \citenamefont {Schroeder}, \citenamefont {Ding} \emph
  {et~al.}}]{borodin2017liquid}%
  \BibitemOpen
  \bibfield  {author} {\bibinfo {author} {\bibfnamefont {O.}~\bibnamefont
  {Borodin}}, \bibinfo {author} {\bibfnamefont {L.}~\bibnamefont {Suo}},
  \bibinfo {author} {\bibfnamefont {M.}~\bibnamefont {Gobet}}, \bibinfo
  {author} {\bibfnamefont {X.}~\bibnamefont {Ren}}, \bibinfo {author}
  {\bibfnamefont {F.}~\bibnamefont {Wang}}, \bibinfo {author} {\bibfnamefont
  {A.}~\bibnamefont {Faraone}}, \bibinfo {author} {\bibfnamefont
  {J.}~\bibnamefont {Peng}}, \bibinfo {author} {\bibfnamefont {M.}~\bibnamefont
  {Olguin}}, \bibinfo {author} {\bibfnamefont {M.}~\bibnamefont {Schroeder}},
  \bibinfo {author} {\bibfnamefont {M.~S.}\ \bibnamefont {Ding}},  \emph
  {et~al.},\ }\href@noop {} {\bibfield  {journal} {\bibinfo  {journal} {ACS
  nano}\ }\textbf {\bibinfo {volume} {11}},\ \bibinfo {pages} {10462} (\bibinfo
  {year} {2017})}\BibitemShut {NoStop}%
\bibitem [{\citenamefont {Choi}\ \emph {et~al.}(2018)\citenamefont {Choi},
  \citenamefont {Lee}, \citenamefont {Choi},\ and\ \citenamefont
  {Cho}}]{choi2018graph}%
  \BibitemOpen
  \bibfield  {author} {\bibinfo {author} {\bibfnamefont {J.-H.}\ \bibnamefont
  {Choi}}, \bibinfo {author} {\bibfnamefont {H.}~\bibnamefont {Lee}}, \bibinfo
  {author} {\bibfnamefont {H.~R.}\ \bibnamefont {Choi}}, \ and\ \bibinfo
  {author} {\bibfnamefont {M.}~\bibnamefont {Cho}},\ }\href@noop {} {\bibfield
  {journal} {\bibinfo  {journal} {Annu. Rev. Phys. Chem.}\ }\textbf {\bibinfo
  {volume} {69}},\ \bibinfo {pages} {125} (\bibinfo {year} {2018})}\BibitemShut
  {NoStop}%
\bibitem [{\citenamefont {Jeon}\ \emph {et~al.}(2020)\citenamefont {Jeon},
  \citenamefont {Lee}, \citenamefont {Choi},\ and\ \citenamefont
  {Cho}}]{jeon2020modeling}%
  \BibitemOpen
  \bibfield  {author} {\bibinfo {author} {\bibfnamefont {J.}~\bibnamefont
  {Jeon}}, \bibinfo {author} {\bibfnamefont {H.}~\bibnamefont {Lee}}, \bibinfo
  {author} {\bibfnamefont {J.-H.}\ \bibnamefont {Choi}}, \ and\ \bibinfo
  {author} {\bibfnamefont {M.}~\bibnamefont {Cho}},\ }\href@noop {} {\bibfield
  {journal} {\bibinfo  {journal} {J. Phys. Chem. C}\ } (\bibinfo {year}
  {2020})}\BibitemShut {NoStop}%
\bibitem [{\citenamefont {McEldrew}\ \emph {et~al.}(2020)\citenamefont
  {McEldrew}, \citenamefont {Goodwin}, \citenamefont {Bi}, \citenamefont
  {Bazant},\ and\ \citenamefont {Kornyshev}}]{mceldrew2020theory}%
  \BibitemOpen
  \bibfield  {author} {\bibinfo {author} {\bibfnamefont {M.}~\bibnamefont
  {McEldrew}}, \bibinfo {author} {\bibfnamefont {Z.~A.}\ \bibnamefont
  {Goodwin}}, \bibinfo {author} {\bibfnamefont {S.}~\bibnamefont {Bi}},
  \bibinfo {author} {\bibfnamefont {M.~Z.}\ \bibnamefont {Bazant}}, \ and\
  \bibinfo {author} {\bibfnamefont {A.~A.}\ \bibnamefont {Kornyshev}},\
  }\href@noop {} {\bibfield  {journal} {\bibinfo  {journal} {J. Chem. Phys.}\
  }\textbf {\bibinfo {volume} {152}},\ \bibinfo {pages} {234506} (\bibinfo
  {year} {2020})}\BibitemShut {NoStop}%
\bibitem [{\citenamefont {McEldrew}\ \emph
  {et~al.}(2021{\natexlab{a}})\citenamefont {McEldrew}, \citenamefont
  {Goodwin}, \citenamefont {Zhao}, \citenamefont {Bazant},\ and\ \citenamefont
  {Kornyshev}}]{mceldrew2020correlated}%
  \BibitemOpen
  \bibfield  {author} {\bibinfo {author} {\bibfnamefont {M.}~\bibnamefont
  {McEldrew}}, \bibinfo {author} {\bibfnamefont {Z.~A.~H.}\ \bibnamefont
  {Goodwin}}, \bibinfo {author} {\bibfnamefont {H.}~\bibnamefont {Zhao}},
  \bibinfo {author} {\bibfnamefont {M.~Z.}\ \bibnamefont {Bazant}}, \ and\
  \bibinfo {author} {\bibfnamefont {A.~A.}\ \bibnamefont {Kornyshev}},\
  }\href@noop {} {\bibfield  {journal} {\bibinfo  {journal} {J. Phys. Chem. B}\
  }\textbf {\bibinfo {volume} {125}},\ \bibinfo {pages} {2677} (\bibinfo {year}
  {2021}{\natexlab{a}})}\BibitemShut {NoStop}%
\bibitem [{\citenamefont {McEldrew}\ \emph
  {et~al.}(2021{\natexlab{b}})\citenamefont {McEldrew}, \citenamefont
  {Goodwin}, \citenamefont {Bi}, \citenamefont {Korvnyshev},\ and\
  \citenamefont {Bazant}}]{mceldrew2021wise}%
  \BibitemOpen
  \bibfield  {author} {\bibinfo {author} {\bibfnamefont {M.}~\bibnamefont
  {McEldrew}}, \bibinfo {author} {\bibfnamefont {Z.~A.}\ \bibnamefont
  {Goodwin}}, \bibinfo {author} {\bibfnamefont {S.}~\bibnamefont {Bi}},
  \bibinfo {author} {\bibfnamefont {A.}~\bibnamefont {Korvnyshev}}, \ and\
  \bibinfo {author} {\bibfnamefont {M.~Z.}\ \bibnamefont {Bazant}},\
  }\href@noop {} {\bibfield  {journal} {\bibinfo  {journal} {J. Electrochem.
  Soc.}\ }\textbf {\bibinfo {volume} {168}},\ \bibinfo {pages} {050514}
  (\bibinfo {year} {2021}{\natexlab{b}})}\BibitemShut {NoStop}%
\bibitem [{\citenamefont {McEldrew}\ \emph
  {et~al.}(2021{\natexlab{c}})\citenamefont {McEldrew}, \citenamefont
  {Goodwin}, \citenamefont {Molinari}, \citenamefont {Kozinsky}, \citenamefont
  {Kornyshev},\ and\ \citenamefont {Bazant}}]{mceldrew2021salt}%
  \BibitemOpen
  \bibfield  {author} {\bibinfo {author} {\bibfnamefont {M.}~\bibnamefont
  {McEldrew}}, \bibinfo {author} {\bibfnamefont {Z.~A.}\ \bibnamefont
  {Goodwin}}, \bibinfo {author} {\bibfnamefont {N.}~\bibnamefont {Molinari}},
  \bibinfo {author} {\bibfnamefont {B.}~\bibnamefont {Kozinsky}}, \bibinfo
  {author} {\bibfnamefont {A.~A.}\ \bibnamefont {Kornyshev}}, \ and\ \bibinfo
  {author} {\bibfnamefont {M.~Z.}\ \bibnamefont {Bazant}},\ }\href@noop {}
  {\bibfield  {journal} {\bibinfo  {journal} {J. Phys. Chem. B}\ }\textbf
  {\bibinfo {volume} {125}},\ \bibinfo {pages} {13752} (\bibinfo {year}
  {2021}{\natexlab{c}})}\BibitemShut {NoStop}%
\bibitem [{\citenamefont {Flory}(1941{\natexlab{a}})}]{flory1941molecular}%
  \BibitemOpen
  \bibfield  {author} {\bibinfo {author} {\bibfnamefont {P.~J.}\ \bibnamefont
  {Flory}},\ }\href@noop {} {\bibfield  {journal} {\bibinfo  {journal} {Journal
  of the American Chemical Society}\ }\textbf {\bibinfo {volume} {63}},\
  \bibinfo {pages} {3083} (\bibinfo {year} {1941}{\natexlab{a}})}\BibitemShut
  {NoStop}%
\bibitem [{\citenamefont {Flory}(1941{\natexlab{b}})}]{flory1941molecular2}%
  \BibitemOpen
  \bibfield  {author} {\bibinfo {author} {\bibfnamefont {P.~J.}\ \bibnamefont
  {Flory}},\ }\href@noop {} {\bibfield  {journal} {\bibinfo  {journal} {Journal
  of the American Chemical Society}\ }\textbf {\bibinfo {volume} {63}},\
  \bibinfo {pages} {3091} (\bibinfo {year} {1941}{\natexlab{b}})}\BibitemShut
  {NoStop}%
\bibitem [{\citenamefont
  {Flory}(1942{\natexlab{a}})}]{flory1942thermodynamics}%
  \BibitemOpen
  \bibfield  {author} {\bibinfo {author} {\bibfnamefont {P.~J.}\ \bibnamefont
  {Flory}},\ }\href@noop {} {\bibfield  {journal} {\bibinfo  {journal} {J.
  Chem. Phys.}\ }\textbf {\bibinfo {volume} {10}},\ \bibinfo {pages} {51}
  (\bibinfo {year} {1942}{\natexlab{a}})}\BibitemShut {NoStop}%
\bibitem [{\citenamefont {Flory}(1942{\natexlab{b}})}]{flory1942constitution}%
  \BibitemOpen
  \bibfield  {author} {\bibinfo {author} {\bibfnamefont {P.~J.}\ \bibnamefont
  {Flory}},\ }\href@noop {} {\bibfield  {journal} {\bibinfo  {journal} {The
  Journal of Physical Chemistry}\ }\textbf {\bibinfo {volume} {46}},\ \bibinfo
  {pages} {132} (\bibinfo {year} {1942}{\natexlab{b}})}\BibitemShut {NoStop}%
\bibitem [{\citenamefont {Flory}(1953)}]{flory1953principles}%
  \BibitemOpen
  \bibfield  {author} {\bibinfo {author} {\bibfnamefont {P.~J.}\ \bibnamefont
  {Flory}},\ }\href@noop {} {\emph {\bibinfo {title} {Principles of polymer
  chemistry}}}\ (\bibinfo  {publisher} {Cornell University Press},\ \bibinfo
  {year} {1953})\BibitemShut {NoStop}%
\bibitem [{\citenamefont {Stockmayer}(1943)}]{stockmayer1943theory}%
  \BibitemOpen
  \bibfield  {author} {\bibinfo {author} {\bibfnamefont {W.~H.}\ \bibnamefont
  {Stockmayer}},\ }\href@noop {} {\bibfield  {journal} {\bibinfo  {journal} {J.
  Chem. Phys.}\ }\textbf {\bibinfo {volume} {11}},\ \bibinfo {pages} {45}
  (\bibinfo {year} {1943})}\BibitemShut {NoStop}%
\bibitem [{\citenamefont {Stockmayer}(1944)}]{stockmayer1944theory}%
  \BibitemOpen
  \bibfield  {author} {\bibinfo {author} {\bibfnamefont {W.~H.}\ \bibnamefont
  {Stockmayer}},\ }\href@noop {} {\bibfield  {journal} {\bibinfo  {journal} {J.
  Chem. Phys.}\ }\textbf {\bibinfo {volume} {12}},\ \bibinfo {pages} {125}
  (\bibinfo {year} {1944})}\BibitemShut {NoStop}%
\bibitem [{\citenamefont {Tanaka}(1989)}]{tanaka1989}%
  \BibitemOpen
  \bibfield  {author} {\bibinfo {author} {\bibfnamefont {F.}~\bibnamefont
  {Tanaka}},\ }\href@noop {} {\bibfield  {journal} {\bibinfo  {journal}
  {Macromolecules}\ }\textbf {\bibinfo {volume} {22}},\ \bibinfo {pages} {1988}
  (\bibinfo {year} {1989})}\BibitemShut {NoStop}%
\bibitem [{\citenamefont {Tanaka}(1990)}]{tanaka1990thermodynamic}%
  \BibitemOpen
  \bibfield  {author} {\bibinfo {author} {\bibfnamefont {F.}~\bibnamefont
  {Tanaka}},\ }\href@noop {} {\bibfield  {journal} {\bibinfo  {journal}
  {Macromolecules}\ }\textbf {\bibinfo {volume} {23}},\ \bibinfo {pages} {3784}
  (\bibinfo {year} {1990})}\BibitemShut {NoStop}%
\bibitem [{\citenamefont {Tanaka}\ and\ \citenamefont
  {Stockmayer}(1994)}]{tanaka1994}%
  \BibitemOpen
  \bibfield  {author} {\bibinfo {author} {\bibfnamefont {F.}~\bibnamefont
  {Tanaka}}\ and\ \bibinfo {author} {\bibfnamefont {W.~H.}\ \bibnamefont
  {Stockmayer}},\ }\href@noop {} {\bibfield  {journal} {\bibinfo  {journal}
  {Macromolecules}\ }\textbf {\bibinfo {volume} {27}},\ \bibinfo {pages} {3943}
  (\bibinfo {year} {1994})}\BibitemShut {NoStop}%
\bibitem [{\citenamefont {Tanaka}\ and\ \citenamefont
  {Ishida}(1995)}]{tanaka1995}%
  \BibitemOpen
  \bibfield  {author} {\bibinfo {author} {\bibfnamefont {F.}~\bibnamefont
  {Tanaka}}\ and\ \bibinfo {author} {\bibfnamefont {M.}~\bibnamefont
  {Ishida}},\ }\href@noop {} {\bibfield  {journal} {\bibinfo  {journal} {J.
  Chem. Soc. Faraday Trans.}\ }\textbf {\bibinfo {volume} {91}},\ \bibinfo
  {pages} {2663} (\bibinfo {year} {1995})}\BibitemShut {NoStop}%
\bibitem [{\citenamefont {Ishida}\ and\ \citenamefont
  {Tanaka}(1997)}]{ishida1997}%
  \BibitemOpen
  \bibfield  {author} {\bibinfo {author} {\bibfnamefont {M.}~\bibnamefont
  {Ishida}}\ and\ \bibinfo {author} {\bibfnamefont {F.}~\bibnamefont
  {Tanaka}},\ }\href@noop {} {\bibfield  {journal} {\bibinfo  {journal}
  {Macromolecules}\ }\textbf {\bibinfo {volume} {30}},\ \bibinfo {pages} {3900}
  (\bibinfo {year} {1997})}\BibitemShut {NoStop}%
\bibitem [{\citenamefont {Tanaka}(1998)}]{tanaka1998}%
  \BibitemOpen
  \bibfield  {author} {\bibinfo {author} {\bibfnamefont {F.}~\bibnamefont
  {Tanaka}},\ }\href@noop {} {\bibfield  {journal} {\bibinfo  {journal}
  {Physica A: Statistical Mechanics and its Applications}\ }\textbf {\bibinfo
  {volume} {257}},\ \bibinfo {pages} {245} (\bibinfo {year}
  {1998})}\BibitemShut {NoStop}%
\bibitem [{\citenamefont {Tanaka}\ and\ \citenamefont
  {Ishida}(1999)}]{tanaka1999}%
  \BibitemOpen
  \bibfield  {author} {\bibinfo {author} {\bibfnamefont {F.}~\bibnamefont
  {Tanaka}}\ and\ \bibinfo {author} {\bibfnamefont {M.}~\bibnamefont
  {Ishida}},\ }\href@noop {} {\bibfield  {journal} {\bibinfo  {journal}
  {Macromolecules}\ }\textbf {\bibinfo {volume} {32}},\ \bibinfo {pages} {1271}
  (\bibinfo {year} {1999})}\BibitemShut {NoStop}%
\bibitem [{\citenamefont {Tanaka}(2002)}]{tanaka2002}%
  \BibitemOpen
  \bibfield  {author} {\bibinfo {author} {\bibfnamefont {F.}~\bibnamefont
  {Tanaka}},\ }\href@noop {} {\bibfield  {journal} {\bibinfo  {journal} {Polym.
  J.}\ }\textbf {\bibinfo {volume} {34}},\ \bibinfo {pages} {479} (\bibinfo
  {year} {2002})}\BibitemShut {NoStop}%
\bibitem [{\citenamefont {Reber}\ \emph {et~al.}(2020)\citenamefont {Reber},
  \citenamefont {Grissa}, \citenamefont {Becker}, \citenamefont {K\"{u}hnel},\
  and\ \citenamefont {Battaglia}}]{Reber2020}%
  \BibitemOpen
  \bibfield  {author} {\bibinfo {author} {\bibfnamefont {D.}~\bibnamefont
  {Reber}}, \bibinfo {author} {\bibfnamefont {R.}~\bibnamefont {Grissa}},
  \bibinfo {author} {\bibfnamefont {M.}~\bibnamefont {Becker}}, \bibinfo
  {author} {\bibfnamefont {R.-S.}\ \bibnamefont {K\"{u}hnel}}, \ and\ \bibinfo
  {author} {\bibfnamefont {C.}~\bibnamefont {Battaglia}},\ }\href@noop {}
  {\bibfield  {journal} {\bibinfo  {journal} {Adv. Energy Mater.}\ }\textbf
  {\bibinfo {volume} {11}},\ \bibinfo {pages} {2002913} (\bibinfo {year}
  {2020})}\BibitemShut {NoStop}%
\bibitem [{\citenamefont {France-Lanord}\ and\ \citenamefont
  {Grossman}(2019)}]{france2019}%
  \BibitemOpen
  \bibfield  {author} {\bibinfo {author} {\bibfnamefont {A.}~\bibnamefont
  {France-Lanord}}\ and\ \bibinfo {author} {\bibfnamefont {J.~C.}\ \bibnamefont
  {Grossman}},\ }\href@noop {} {\bibfield  {journal} {\bibinfo  {journal}
  {Phys. Rev. Lett.}\ }\textbf {\bibinfo {volume} {122}},\ \bibinfo {pages}
  {136001} (\bibinfo {year} {2019})}\BibitemShut {NoStop}%
\bibitem [{\citenamefont {Goodwin}\ \emph {et~al.}(2022)\citenamefont
  {Goodwin}, \citenamefont {McEldrew}, \citenamefont {de~Souza}, \citenamefont
  {Bazant},\ and\ \citenamefont {Kornyshev}}]{Goodwin2022EDLgel}%
  \BibitemOpen
  \bibfield  {author} {\bibinfo {author} {\bibfnamefont {Z.~A.~H.}\
  \bibnamefont {Goodwin}}, \bibinfo {author} {\bibfnamefont {M.}~\bibnamefont
  {McEldrew}}, \bibinfo {author} {\bibfnamefont {J.~P.}\ \bibnamefont
  {de~Souza}}, \bibinfo {author} {\bibfnamefont {M.~Z.}\ \bibnamefont
  {Bazant}}, \ and\ \bibinfo {author} {\bibfnamefont {A.~A.}\ \bibnamefont
  {Kornyshev}},\ }\href@noop {} {\bibfield  {journal} {\bibinfo  {journal}
  {arXiv:2204.11123}\ } (\bibinfo {year} {2022})}\BibitemShut {NoStop}%
\bibitem [{\citenamefont {Budkov}\ \emph {et~al.}(2018)\citenamefont {Budkov},
  \citenamefont {Kolesnikov}, \citenamefont {Goodwin}, \citenamefont
  {Kiselev},\ and\ \citenamefont {Kornyshev}}]{BBKGK}%
  \BibitemOpen
  \bibfield  {author} {\bibinfo {author} {\bibfnamefont {Y.~A.}\ \bibnamefont
  {Budkov}}, \bibinfo {author} {\bibfnamefont {A.~L.}\ \bibnamefont
  {Kolesnikov}}, \bibinfo {author} {\bibfnamefont {Z.~A.~H.}\ \bibnamefont
  {Goodwin}}, \bibinfo {author} {\bibfnamefont {M.}~\bibnamefont {Kiselev}}, \
  and\ \bibinfo {author} {\bibfnamefont {A.~A.}\ \bibnamefont {Kornyshev}},\
  }\href@noop {} {\bibfield  {journal} {\bibinfo  {journal} {Electrochim.
  Acta}\ }\textbf {\bibinfo {volume} {284}},\ \bibinfo {pages} {346} (\bibinfo
  {year} {2018})}\BibitemShut {NoStop}%
\bibitem [{\citenamefont {Zwanikken}\ and\ \citenamefont {van
  Roij}(2009)}]{IBL}%
  \BibitemOpen
  \bibfield  {author} {\bibinfo {author} {\bibfnamefont {J.}~\bibnamefont
  {Zwanikken}}\ and\ \bibinfo {author} {\bibfnamefont {R.}~\bibnamefont {van
  Roij}},\ }\href@noop {} {\bibfield  {journal} {\bibinfo  {journal} {J. Phys.
  Condens. Matter}\ }\textbf {\bibinfo {volume} {41}},\ \bibinfo {pages}
  {424102} (\bibinfo {year} {2009})}\BibitemShut {NoStop}%
\bibitem [{\citenamefont {Kornyshev}(2007)}]{Kornyshev2007}%
  \BibitemOpen
  \bibfield  {author} {\bibinfo {author} {\bibfnamefont {A.~A.}\ \bibnamefont
  {Kornyshev}},\ }\href {\doibase 10.1021/jp067857o} {\bibfield  {journal}
  {\bibinfo  {journal} {J. Phys. Chem. B}\ }\textbf {\bibinfo {volume} {111}},\
  \bibinfo {pages} {5545} (\bibinfo {year} {2007})}\BibitemShut {NoStop}%
\bibitem [{\citenamefont {Kilic}\ \emph {et~al.}(2007)\citenamefont {Kilic},
  \citenamefont {Bazant},\ and\ \citenamefont {Ajdari}}]{kilic2007a}%
  \BibitemOpen
  \bibfield  {author} {\bibinfo {author} {\bibfnamefont {M.~S.}\ \bibnamefont
  {Kilic}}, \bibinfo {author} {\bibfnamefont {M.~Z.}\ \bibnamefont {Bazant}}, \
  and\ \bibinfo {author} {\bibfnamefont {A.}~\bibnamefont {Ajdari}},\
  }\href@noop {} {\bibfield  {journal} {\bibinfo  {journal} {Phys. Rev. E}\
  }\textbf {\bibinfo {volume} {75}},\ \bibinfo {pages} {021502} (\bibinfo
  {year} {2007})}\BibitemShut {NoStop}%
\bibitem [{\citenamefont {Bazant}\ \emph {et~al.}(2009)\citenamefont {Bazant},
  \citenamefont {Kilic}, \citenamefont {Storey},\ and\ \citenamefont
  {Ajdari}}]{Bazant2009a}%
  \BibitemOpen
  \bibfield  {author} {\bibinfo {author} {\bibfnamefont {M.~Z.}\ \bibnamefont
  {Bazant}}, \bibinfo {author} {\bibfnamefont {M.~S.}\ \bibnamefont {Kilic}},
  \bibinfo {author} {\bibfnamefont {B.}~\bibnamefont {Storey}}, \ and\ \bibinfo
  {author} {\bibfnamefont {A.}~\bibnamefont {Ajdari}},\ }\href@noop {}
  {\bibfield  {journal} {\bibinfo  {journal} {Adv. Colloid Interface Sci.}\
  }\textbf {\bibinfo {volume} {152}},\ \bibinfo {pages} {48} (\bibinfo {year}
  {2009})}\BibitemShut {NoStop}%
\bibitem [{\citenamefont {Jitvisate}\ and\ \citenamefont
  {Seddon}(2018)}]{Monchai2018}%
  \BibitemOpen
  \bibfield  {author} {\bibinfo {author} {\bibfnamefont {M.}~\bibnamefont
  {Jitvisate}}\ and\ \bibinfo {author} {\bibfnamefont {J.~R.~T.}\ \bibnamefont
  {Seddon}},\ }\href@noop {} {\bibfield  {journal} {\bibinfo  {journal} {J.
  Phys. Chem. Lett.}\ }\textbf {\bibinfo {volume} {9}},\ \bibinfo {pages} {126}
  (\bibinfo {year} {2018})}\BibitemShut {NoStop}%
\bibitem [{\citenamefont {Onsager}(1934)}]{Onsager1934}%
  \BibitemOpen
  \bibfield  {author} {\bibinfo {author} {\bibfnamefont {L.}~\bibnamefont
  {Onsager}},\ }\href@noop {} {\bibfield  {journal} {\bibinfo  {journal} {J.
  Chem. Phys.}\ }\textbf {\bibinfo {volume} {2}},\ \bibinfo {pages} {599}
  (\bibinfo {year} {1934})}\BibitemShut {NoStop}%
\bibitem [{\citenamefont {Kaiser}\ \emph {et~al.}(2013)\citenamefont {Kaiser},
  \citenamefont {Bramwell}, \citenamefont {Holdsworth},\ and\ \citenamefont
  {Moessner}}]{Kaiser2013}%
  \BibitemOpen
  \bibfield  {author} {\bibinfo {author} {\bibfnamefont {V.}~\bibnamefont
  {Kaiser}}, \bibinfo {author} {\bibfnamefont {S.~T.}\ \bibnamefont
  {Bramwell}}, \bibinfo {author} {\bibfnamefont {P.~C.}\ \bibnamefont
  {Holdsworth}}, \ and\ \bibinfo {author} {\bibfnamefont {R.}~\bibnamefont
  {Moessner}},\ }\href@noop {} {\bibfield  {journal} {\bibinfo  {journal} {Nat.
  Mater}\ }\textbf {\bibinfo {volume} {12}},\ \bibinfo {pages} {1033} (\bibinfo
  {year} {2013})}\BibitemShut {NoStop}%
\bibitem [{\citenamefont {Onsager}\ and\ \citenamefont
  {Kim}(1957)}]{Shoon1957}%
  \BibitemOpen
  \bibfield  {author} {\bibinfo {author} {\bibfnamefont {L.}~\bibnamefont
  {Onsager}}\ and\ \bibinfo {author} {\bibfnamefont {S.~K.}\ \bibnamefont
  {Kim}},\ }\href@noop {} {\bibfield  {journal} {\bibinfo  {journal} {. Phys.
  Chem.}\ }\textbf {\bibinfo {volume} {61}},\ \bibinfo {pages} {198} (\bibinfo
  {year} {1957})}\BibitemShut {NoStop}%
\bibitem [{\citenamefont {Kumar}\ \emph {et~al.}(2015)\citenamefont {Kumar},
  \citenamefont {Bocharova}, \citenamefont {Strelcov}, \citenamefont {Tselev},
  \citenamefont {Kravchenko}, \citenamefont {Berdzinski}, \citenamefont
  {Strehmel}, \citenamefont {Ovchinnikova}, \citenamefont {Minutolo},
  \citenamefont {Sangoro}, \citenamefont {Agapov}, \citenamefont {Sokolov},
  \citenamefont {Kalinin},\ and\ \citenamefont {Sumpter}}]{Kumar2015}%
  \BibitemOpen
  \bibfield  {author} {\bibinfo {author} {\bibfnamefont {R.}~\bibnamefont
  {Kumar}}, \bibinfo {author} {\bibfnamefont {V.}~\bibnamefont {Bocharova}},
  \bibinfo {author} {\bibfnamefont {E.}~\bibnamefont {Strelcov}}, \bibinfo
  {author} {\bibfnamefont {A.}~\bibnamefont {Tselev}}, \bibinfo {author}
  {\bibfnamefont {I.~I.}\ \bibnamefont {Kravchenko}}, \bibinfo {author}
  {\bibfnamefont {S.}~\bibnamefont {Berdzinski}}, \bibinfo {author}
  {\bibfnamefont {V.}~\bibnamefont {Strehmel}}, \bibinfo {author}
  {\bibfnamefont {O.~S.}\ \bibnamefont {Ovchinnikova}}, \bibinfo {author}
  {\bibfnamefont {J.~A.}\ \bibnamefont {Minutolo}}, \bibinfo {author}
  {\bibfnamefont {J.~R.}\ \bibnamefont {Sangoro}}, \bibinfo {author}
  {\bibfnamefont {A.~L.}\ \bibnamefont {Agapov}}, \bibinfo {author}
  {\bibfnamefont {A.~P.}\ \bibnamefont {Sokolov}}, \bibinfo {author}
  {\bibfnamefont {S.~V.}\ \bibnamefont {Kalinin}}, \ and\ \bibinfo {author}
  {\bibfnamefont {B.~G.}\ \bibnamefont {Sumpter}},\ }\href@noop {} {\bibfield
  {journal} {\bibinfo  {journal} {Nanoscale}\ }\textbf {\bibinfo {volume}
  {7}},\ \bibinfo {pages} {947} (\bibinfo {year} {2015})}\BibitemShut {NoStop}%
\bibitem [{\citenamefont {Patro}\ \emph {et~al.}(2016)\citenamefont {Patro},
  \citenamefont {Burghaus},\ and\ \citenamefont {Roling}}]{Patro2016}%
  \BibitemOpen
  \bibfield  {author} {\bibinfo {author} {\bibfnamefont {L.~N.}\ \bibnamefont
  {Patro}}, \bibinfo {author} {\bibfnamefont {O.}~\bibnamefont {Burghaus}}, \
  and\ \bibinfo {author} {\bibfnamefont {B.}~\bibnamefont {Roling}},\
  }\href@noop {} {\bibfield  {journal} {\bibinfo  {journal} {Phys. Rev. Lett.}\
  }\textbf {\bibinfo {volume} {116}},\ \bibinfo {pages} {185901} (\bibinfo
  {year} {2016})}\BibitemShut {NoStop}%
\bibitem [{\citenamefont {Roling}\ \emph {et~al.}(2017)\citenamefont {Roling},
  \citenamefont {Patro}, \citenamefont {Burghaus},\ and\ \citenamefont
  {Gr\"{a}f}}]{Roling2017}%
  \BibitemOpen
  \bibfield  {author} {\bibinfo {author} {\bibfnamefont {B.}~\bibnamefont
  {Roling}}, \bibinfo {author} {\bibfnamefont {L.}~\bibnamefont {Patro}},
  \bibinfo {author} {\bibfnamefont {O.}~\bibnamefont {Burghaus}}, \ and\
  \bibinfo {author} {\bibfnamefont {M.}~\bibnamefont {Gr\"{a}f}},\ }\href@noop
  {} {\bibfield  {journal} {\bibinfo  {journal} {Eur. Phys. J. Special Topics}\
  }\textbf {\bibinfo {volume} {226}},\ \bibinfo {pages} {3095} (\bibinfo {year}
  {2017})}\BibitemShut {NoStop}%
\bibitem [{\citenamefont {MacFarlane}\ \emph {et~al.}(2009)\citenamefont
  {MacFarlane}, \citenamefont {Forsyth}, \citenamefont {Izgorodina},
  \citenamefont {Abbott}, \citenamefont {Annata},\ and\ \citenamefont
  {Fraser}}]{MacFarlane2009}%
  \BibitemOpen
  \bibfield  {author} {\bibinfo {author} {\bibfnamefont {D.~R.}\ \bibnamefont
  {MacFarlane}}, \bibinfo {author} {\bibfnamefont {M.}~\bibnamefont {Forsyth}},
  \bibinfo {author} {\bibfnamefont {E.~I.}\ \bibnamefont {Izgorodina}},
  \bibinfo {author} {\bibfnamefont {A.~P.}\ \bibnamefont {Abbott}}, \bibinfo
  {author} {\bibfnamefont {G.}~\bibnamefont {Annata}}, \ and\ \bibinfo {author}
  {\bibfnamefont {K.}~\bibnamefont {Fraser}},\ }\href@noop {} {\bibfield
  {journal} {\bibinfo  {journal} {Phys. Chem. Chem. Phys.}\ }\textbf {\bibinfo
  {volume} {11}},\ \bibinfo {pages} {4962} (\bibinfo {year}
  {2009})}\BibitemShut {NoStop}%
\bibitem [{\citenamefont {Downing}\ \emph {et~al.}(2018)\citenamefont
  {Downing}, \citenamefont {Bossa},\ and\ \citenamefont {May}}]{Downing2018}%
  \BibitemOpen
  \bibfield  {author} {\bibinfo {author} {\bibfnamefont {R.}~\bibnamefont
  {Downing}}, \bibinfo {author} {\bibfnamefont {G.~V.}\ \bibnamefont {Bossa}},
  \ and\ \bibinfo {author} {\bibfnamefont {S.}~\bibnamefont {May}},\
  }\href@noop {} {\bibfield  {journal} {\bibinfo  {journal} {J. Phys. Chem. C}\
  }\textbf {\bibinfo {volume} {122}},\ \bibinfo {pages} {28537} (\bibinfo
  {year} {2018})}\BibitemShut {NoStop}%
\bibitem [{\citenamefont {Bossa}\ \emph {et~al.}(2015)\citenamefont {Bossa},
  \citenamefont {Roth},\ and\ \citenamefont {May}}]{Bossa2015}%
  \BibitemOpen
  \bibfield  {author} {\bibinfo {author} {\bibfnamefont {G.~V.}\ \bibnamefont
  {Bossa}}, \bibinfo {author} {\bibfnamefont {J.}~\bibnamefont {Roth}}, \ and\
  \bibinfo {author} {\bibfnamefont {S.}~\bibnamefont {May}},\ }\href@noop {}
  {\bibfield  {journal} {\bibinfo  {journal} {Langmuir}\ }\textbf {\bibinfo
  {volume} {31}},\ \bibinfo {pages} {9924} (\bibinfo {year}
  {2015})}\BibitemShut {NoStop}%
\bibitem [{\citenamefont {Marcus}\ and\ \citenamefont
  {Hefter}(2006)}]{marcus2006ion}%
  \BibitemOpen
  \bibfield  {author} {\bibinfo {author} {\bibfnamefont {Y.}~\bibnamefont
  {Marcus}}\ and\ \bibinfo {author} {\bibfnamefont {G.}~\bibnamefont
  {Hefter}},\ }\href@noop {} {\bibfield  {journal} {\bibinfo  {journal}
  {Chemical reviews}\ }\textbf {\bibinfo {volume} {106}},\ \bibinfo {pages}
  {4585} (\bibinfo {year} {2006})}\BibitemShut {NoStop}%
\end{thebibliography}%

\end{document}